\documentclass[aps,prd,preprint,superscriptaddress,nofootinbib,amsmath,amssymb]{revtex4}
\pdfoutput=1
\usepackage{graphicx}
\usepackage{dcolumn}
\usepackage{bm}
\usepackage{amsmath}
\usepackage{amsfonts}
\usepackage{amssymb}
\usepackage{appendix}
\usepackage{pdfpages}
\usepackage{graphicx}
\usepackage{wrapfig}
\usepackage{multirow}
\usepackage{mathrsfs}
\usepackage{epstopdf}
\usepackage{slashed}
\usepackage{soul}
\usepackage{mathtools}
\usepackage{floatrow}
\usepackage{float}
\usepackage{csquotes}
\usepackage{wrapfig}
\usepackage{graphicx}
\usepackage{epsfig}
\usepackage[utf8]{inputenc}
\usepackage{epsfig}
\usepackage{dcolumn}
\usepackage{morefloats}
\usepackage{hyperref}
\newcommand{\ba}{\begin{array}}
\newcommand{\ea}{\end{array}}
\def\be{\begin{equation}}
\def\ee{\end{equation}}
\def\bea{\begin{eqnarray}}
\def\eea{\end{eqnarray}}
\def\gsim{\ \rlap{\raise 2pt\hbox{$>$}}{\lower 2pt \hbox{$\sim$}}\ }
\def\lsim{\ \rlap{\raise 2pt\hbox{$<$}}{\lower 2pt \hbox{$\sim$}}\ }
\def\dslash{\kern-4pt \not{\hbox{\kern-2pt $\partial$}}}
\def\pslash{\not{\hbox{\kern-2pt p}}}

\begin{document}
\DeclareGraphicsExtensions{.eps,.ps}


\title{$ \mu-\tau $ Reflection Symmetry Embedded in Minimal Seesaw}



\author{Newton Nath}
\email[Email Address: ]{newton@ihep.ac.cn}
\affiliation{
Institute of High Energy Physics, Chinese Academy of Sciences, Beijing, 100049, China}
\affiliation{
School of Physical Sciences, University of Chinese Academy of Sciences, Beijing, 100049, China}
\author{Zhi-zhong Xing}
\email[Email Address: ]{xingzz@ihep.ac.cn}
\affiliation{
Institute of High Energy Physics, Chinese Academy of Sciences, Beijing, 100049, China}
\affiliation{
School of Physical Sciences, University of Chinese Academy of Sciences, Beijing, 100049, China}
\affiliation{
Center for High Energy Physics, Peking University, Beijing, 100871, China}
\author{Jue Zhang}
\email[Email Address: ]{juezhang87@pku.edu.cn}
\affiliation{
Center for High Energy Physics, Peking University, Beijing, 100871, China}
\affiliation{Department of Physics and Institute of Theoretical Physics, Nanjing Normal University, Nanjing, 210023, China}

\begin{abstract}
We embed $\mu-\tau$ reflection symmetry into the minimal seesaw formalism, where two right-handed neutrinos are added to the Standard Model of particle physics. Assuming that both the left- and right-handed neutrino fields transform under $\mu-\tau$ reflection symmetry, we obtain the required forms of the neutrino Dirac mass matrix and the Majorana mass matrix for the right-handed neutrinos. To investigate the neutrino phenomenology at low energies, we first consider the breaking of $\mu-\tau$ reflection symmetry due to the renormalization group running, and then systematically study various breaking schemes by introducing explicit breaking terms at high energies. 
\end{abstract}
\maketitle

\section{Introduction}
Over past two decades, phenomenal neutrino oscillation experiments have established the formalism of three flavor neutrino oscillations and determined two mass squared differences and three mixing angles. At present, unknowns in neutrino oscillation physics are: the neutrino mass hierarchy, i.e.,  whether neutrinos obey normal hierarchy (NH, $m_1^{} <  m_2^{} <  m_3^{}$ with $m_i^{}$'s being neutrino masses) or inverted hierarchy (IH, $m_3^{} < m_1^{} \sim m_2^{}$);  the octant of atmospheric mixing angle $\theta_{23}$, 
and the determination of Dirac CP-violating phase $\delta$.\footnote{Current experimental data tend to favor $\delta \sim -90^\circ$~\cite{Capozzi:2016rtj,Esteban:2016qun,deSalas:2017kay}.} Were neutrinos the Majorana particles, there would then exist two additional Majorana phases, which do not affect neutrino oscillation probabilities but can be probed by the neutrinoless double beta-decay ($0\nu\beta\beta $) experiments~\cite{Pas:2015eia}.

In the Standard Model (SM) of particle physics neutrinos are massless. One economical way to incorporate non-zero neutrino masses is to add two right-handed neutrinos to the SM and allow lepton number violation,\footnote{At least two right-handed neutrinos are required to explain the observed two mass squared differences of three active neutrinos. }  resulting in the so-called minimal seesaw scenario (see Ref.~\cite{Guo:2006qa} for a review) within the context of the Type-I seesaw mechanism~\cite{Minkowski:1977sc,Yanagida:1979as,GellMann:1980vs,Mohapatra:1979ia,Schechter:1980gr}. Integrating out the heavy right-handed neutrino fields results in the light neutrino mass matrix $M_\nu^{}$ as $  M_{\nu}^{} \approx - M_D^{} M^{-1}_R M^{T}_D $. In this minimal seesaw set-up, $M_D^{}$  is  the $(3 \times 2)$ neutrino Dirac mass matrix whereas $M_R^{}$ is the ($2 \times 2)$ Majorana mass matrix for the right-handed neutrinos. In the basis where the charged lepton Yukawa matrix $Y_l^{}$ is diagonal, diagonalizing the light neutrino mass matrix $M_\nu^{}$ then leads to the lepton mixing matrix, which is found to be sharply different from the quark mixing matrix. Namely, the former is highly non-diagonal while the latter almost diagonal. To explain the peculiar patterns in the lepton mixing matrix, various flavor symmetry models have been considered, e.g., in Refs.~\cite{Altarelli:2010gt,Altarelli:2012ss,Smirnov:2011jv,Ishimori:2010au,King:2013eh}. 

In this work we focus on the so-called $\mu-\tau$ reflection symmetry, firstly proposed among the left-handed neutrino fields in Ref.~\cite{Harrison:2002et}. Specifically, one imposes the following transformations on the left-handed neutrino fields,
\begin{equation}
\nu_{L, e}^{} \leftrightarrow \nu^{c}_{L, e}, ~~~ \nu_{L, \mu}^{} \leftrightarrow \nu^{c}_{L, \tau}, ~~~\nu_{L, \tau}^{} \leftrightarrow \nu^{c}_{L, \mu} \; ,
\end{equation}
where $\nu_{L,\alpha}^{}$'s (for $\alpha = e, \mu, \tau$) are the left-handed neutrino fields in the flavor basis, and $\nu_{L,\alpha}^c$'s are the corresponding charge-conjugated fields. Such transformations then lead to four relations among the entries of the light neutrino mass matrix $M_\nu^{}$, i.e.,
\begin{eqnarray} \label{eq:Mnu_pred}
(M_\nu^{})_{ee}^{} = (M_\nu^{})_{ee}^* \; , \quad (M_\nu^{})_{\mu\tau}^{} = (M_\nu^{})_{\mu\tau}^* \; , \quad (M_\nu^{})_{e\mu}^{} = (M_\nu^{})_{e\tau}^* \; , \quad (M_\nu^{})_{\mu\mu}^{} = (M_\nu^{})_{\tau\tau}^* \; ,
\end{eqnarray}
and therefore yield the predictions: the maximal atmospheric mixing angle $\theta_{23}^{}$, i.e., $\theta_{23}^{} = 45^\circ$; the values of $\pm 90^\circ$ for the Dirac phase $\delta$; and trivial values for the Majorana phases with non-zero 1-3 mixing angle, $\theta_{13}^{}$. As the lepton mixing angles $\theta_{12}^{}$ and $\theta_{13}^{}$ are not specified by the symmetry, the $\mu-\tau$ reflection symmetry is compatible with current experimental data, and thus recently receives a lot of attention, e.g., see Refs.~\cite{Ferreira:2012ri, Grimus:2012hu, Mohapatra:2012tb, Ma:2015gka,Ma:2015pma,Ma:2015fpa,He:2015xha, Joshipura:2015dsa,
Joshipura:2015zla, Joshipura:2016hvn, Nishi:2016wki,
Zhao:2017yvw,Rodejohann:2017lre,Liu:2017frs,Xing:2017mkx,Xing:2017cwb} and Ref.~\cite{Xing:2015fdg} for the latest review. We note that in the literature there exists a similar but different $\mu-\tau$ flavor symmetry, i.e., the $\mu-\tau$ permutation symmetry~\cite{Fukuyama:1997ky,Ma:2001mr,Lam:2001fb,Balaji:2001ex,Grimus:2004cc,Xing:2012ej,Liao:2012xm,Gupta:2013it,Gomez-Izquierdo:2017med}, which predicts zero $\theta_{13}^{}$.

Other than assigning the $\mu-\tau$ reflection symmetry only to the left-handed neutrino fields, in this work we apply the same symmetry to the right-handed neutrino fields as well. Consequently, both the neutrino Dirac mass matrix $M_D^{}$ and the Majorana mass matrix $M_R^{}$ need to satisfy certain relations among their entries. While the resultant light neutrino mass matrix $M_\nu^{}$ still obeys the relations given in Eq.~(\ref{eq:Mnu_pred}), the $\mu-\tau$ reflection symmetry is now embedded in the minimal seesaw formalism, and both the left- and right-handed neutrinos are treated on the same footing. Similar ideas have been studied for the $\mu-\tau$ permutation symmetry~\cite{Mohapatra:2004hta,Joshipura:2009fu}, while for the $\mu-\tau$ reflection symmetry a detailed study on the scenario as ours is still missing.\footnote{Recently, there also exist several works of discussing the $\mu-\tau$ flavor symmetry in the minimal seesaw setup~\cite{Liu:2017frs,Shimizu:2017fgu,Shimizu:2017vwi,Samanta:2017kce}, however, the considered forms of $M_D^{}$ and $M_R^{}$ are different from ours. } Some recent studies on the minimal seesaw model can be found in Refs.\cite{King:1998jw,King:1999cm,Branco:2002ie,Frampton:2002qc,Bhattacharya:2006aw,GomezIzquierdo:2007vn,Goswami:2008rt,Ge:2010js,Goswami:2008uv,Rodejohann:2012jz,Harigaya:2012bw,Zhang:2015tea,Bambhaniya:2016rbb,Rink:2016knw,Rink:2016vvl}.

In this paper we investigate the implications of the above embedding on neutrino phenomenology at low energies, while the possible discussions on the ultraviolet (UV) aspects, such as explaining the baryon asymmetry via leptogenesis~\cite{Fukugita:1986hr} deferred to the future work. For the low energy neutrino phenomenology, since the resultant light neutrino mass matrix $M_\nu^{}$ in Eq.~(\ref{eq:refle_mat}) still preserves the usual $\mu-\tau$ reflection symmetry, one may conclude that there is no new prediction in this seesaw embedded setup. However, we want to point out here at least two deserving issues which need careful scrutiny. The first one is to study the breaking of $\mu-\tau$ reflection symmetry due to the renormalization group (RG) running. As now we impose the $\mu-\tau$ reflection symmetry above the seesaw mass thresholds, an investigation on the RG-running effects is then inevitable in order to confront with the current global-fit of neutrino oscillation data~\cite{Capozzi:2016rtj,Esteban:2016qun,deSalas:2017kay} at low energies.  Secondly, although the latest experimental data of T2K~\cite{Abe:2017uxa} and NO$\nu$A~\cite{NOVA2018} are in good agreement with the predictions of the exact $\mu-\tau$ reflection symmetry,\footnote{We note that the rejection of maximal mixing in $\theta_{23}^{}$ has moved from $2.6\sigma$~\cite{Adamson:2017gxd} to $0.8\sigma$~\cite{NOVA2018}.} there still exist large uncertainties in the measurement of $\theta_{23}^{}$. For instance, the best-fit values of $\theta_{23}^{}$ for the lower and higher octants are $43.6^\circ$ and $48.3^\circ$ in Ref.~\cite{NOVA2018}, respectively. Thus, it is tenacious to believe the exactness of $ \mu-\tau $ reflection symmetry, especially that there may exist large discrepancies when upcoming experimental data will be included. Therefore, it is also worthwhile to study how we can perturb such an exact $\mu-\tau$ reflection symmetry, so that remarkable deviations can be observed in the lepton mixing parameters.

We organize our paper as follows. In Section~\ref{ses:symm}, we introduce the $\mu-\tau$ reflection symmetry transformations to the left- and right-handed neutrino fields, and discuss the required forms of $M_D^{}$ and $M_R^{}$. In Section~\ref{sec:rge}, we proceed to discuss the breaking of $\mu-\tau$ reflection symmetry due to the RG running, followed by the systematic investigation on all the possible explicit breaking patterns of $M_D^{}$ and $M_R^{}$ in Section~\ref{ses:symm_break}. Finally, we summarize our findings in Section~\ref{ses:conc}. Details of derivations, explanations and numerical results are relegated to the Appendices.

\section{$ \mu-\tau $ reflection symmetry embedded in minimal seesaw} \label{ses:symm}

In the minimal seesaw formalism, two right-handed neutrinos, collectively denoted as $N_R^{} = (N_{\mu R}, N_{\tau R})^T$ in the flavor basis, is added to the SM. The relevant Lagrangian containing the neutrino Yukawa matrix and the Majorana mass term for the right-handed neutrinos are given by
\begin{equation}\label{eq:lag}
-\mathcal{L}\supset  \overline{\ell_{L}}~Y_\nu^{} N_R \widetilde{H} + \dfrac{1}{2}\overline{N^{c}_{R}} M_R N_R + \mathrm{h.c.} \; ,
\end{equation}
where $\ell_L^{}$ is the lepton doublet in the SM, $Y_\nu^{}$ stands for the neutrino Yukawa matrix, and $\widetilde{H} = i \sigma_2^{} H^*$ with $H$ denoting the SM Higgs field. After the Higgs field acquiring its vacuum expectation value, i.e., $v = \langle H \rangle \approx 174~\mathrm{GeV}$, we obtain the neutrino Dirac mass term as $\overline{\nu_L^{}} M_D^{} N_R^{} + \mathrm{h.c.}$, where $M_D^{} = v Y_\nu^{}$ is the neutrino Dirac mass matrix, and $ \nu_L^{} = (\nu_{eL}^{}, \nu_{\mu L}^{}, \nu_{\tau L}^{})^{T}$ stands for the left-handed neutrino fields in the flavor basis. 

To embed the $\mu-\tau $ reflection symmetry into the minimal seesaw formalism, we first propose the following transformations for the left- and right-handed neutrino fields,
\begin{equation} \label{eq:trans}
\nu_L\rightarrow S \nu^{c}_L, \qquad N_R \rightarrow S^{\prime} N^{c}_R
\end{equation}
where $ \nu^{c}_L = C\overline{\nu_L}^{T}$ and $N^{c}_R = C\overline{N_R}^{T} $ are the charge-conjugated fields of $\nu_L^{}$ and $N_R^{}$, respectively, and the transformation matrices $S$ and $S^\prime$ are given by
\begin{equation}
 S=\left(  \ba{cc}
1&0\\
0& S^{\prime}\\ \ea \right) 
~,~~ S^{\prime}=\left(  \ba{cc}
0&1\\
1&0 \\ \ea \right)   \; .
\end{equation} 
Applying the above transformations to the mass terms of neutrinos yields
\begin{eqnarray}
-\mathcal{L} & = & \hspace{-0.2cm}
\overline{\nu^{ c}_{ L}} S M_D S^{\prime} N^{ c}_{ R} + \overline{N^{ c}_{ R}} S^{\prime} M^{\dagger}_D S \nu^{ c}_{ L} +
\dfrac{1}{2}(\overline{N_{R}}S^{\prime} M_R S^{\prime}N^{c}_R + \overline{N^{c}_{R}}S^{\prime} M^{\ast}_R S^{\prime} N_R ), \nonumber \\
\hspace{-0.2cm} 
 & = & \hspace{-0.2cm}
\overline{\nu^{}_{ L}} S M^{\ast}_D S^{\prime} N^{}_{ R}
+ \overline{N^{}_{R}} S^{\prime} M^{ T}_D S \nu^{}_{ L} + 
\dfrac{1}{2}(\overline{N^{c}_{R}}S^{\prime} M^{\ast}_R S^{\prime} N_R  + \overline{N_{R}}S^{\prime} M_R S^{\prime}N^{c}_R )\;.
\end{eqnarray}
Then, if the neutrino mass terms $M_D^{}$ and $M_R^{}$ obey the following relations,
\begin{equation}
M_D=SM^{*}_DS^{\prime}, \qquad M_R=S^{\prime}M^{*}_RS^{\prime} \; ,
\end{equation}
we state that $M_D^{}$ and $M_R^{}$ are invariant under the transformations given in Eq.~(\ref{eq:trans}), or, $\mu-\tau$ reflection symmetry is embedded in both $M_D^{}$ and $M_R^{}$ . 

Without loss of generality, the $\mu-\tau$ reflection symmetric limit of $ M_D $ and $ M_R $ can be parameterized as,
\begin{eqnarray}\label{eq:md}
M_D^{} &=& \begin{pmatrix} b & b^{\ast} \\
 c & d \cr  d^{\ast} & c^{\ast}
\end{pmatrix} = \begin{pmatrix} |b| e^{i \phi_{b}} &  |b| e^{-i \phi_{b}} \\
  |c| e^{i \phi_{c}} &  |d| e^{i \phi_{d}} \cr   |d| e^{-i \phi_{d}} &  |c| e^{-i \phi_{c}}
\end{pmatrix}  \; , \\
\label{eq:mr}
M_R &=& \begin{pmatrix} m_{22} & m_{23} \\
   m_{23} & m^{\ast}_{22}
\end{pmatrix} =  \begin{pmatrix}|m_{22}| e^{i \phi_{m}}  & m_{23} \\
   m_{23} & |m_{22}| e^{-i \phi_{m}} 
\end{pmatrix} \; ,
\end{eqnarray}
where the phases and $m_{23}$ are all real. According to the seesaw mass formula, we then obtain the mass matrix $M_\nu^{}$ for the light neutrinos as,
\begin{eqnarray}\label{eq:refle_mat}
-M_{\nu} = M_D M^{-1}_R M^{T}_D =  \left( \begin{matrix}A & B & B^{*} \cr
  B & C & D \cr
  B^{*} & D & C^{\ast} \cr
\end{matrix} \right), 
\end{eqnarray}
with the parameters $A, B, C$ and $D$ given by
\begin{eqnarray}
A & = &  2 |b|^2 \left[ - m^{\prime}_{23} + m^{\prime}_{22} \cos (2\phi_{b} - \phi_{m}) \right] ,  \nonumber \\
B & = & |b|\left[ (|c|e^{i(\phi_b + \phi_c-\phi_m)}  + |d|e^{i(\phi_d - \phi_b + \phi_m)} )  m^{\prime}_{22} -  (|c|e^{i(\phi_c - \phi_b)} + |d|e^{i(\phi_b + \phi_d)})  m^{\prime}_{23}   \right] \;,  \nonumber \\
C & = & \left( |c|^{2}e^{i(2\phi_c-\phi_m)}  + |d|^2e^{i(2\phi_d + \phi_m)} \right)  m^{\prime}_{22} -  2|c||d|  m^{\prime}_{23} e^{i(\phi_c + \phi_d)} \;,  \nonumber \\
D & = & -( |c|^{2}+|d|^2)  m^{\prime}_{23} + 2 |c||d|  m^{\prime}_{22}\cos(\phi_c-\phi_d-\phi_m) \; .
\end{eqnarray}
Here $ m^{\prime}_{23} = m_{23}/(|m_{22}|^{2}- m^{2}_{23} )  $ and $ m^{\prime}_{22} = |m_{22}|/(|m_{22}|^{2}- m^{2}_{23} )  $. We note that the parameters $A$ and $D$ in $M_\nu^{}$ are real, and $M_\nu^{}$ preserves the usual $\mu-\tau$ reflection symmetry in Eq.~(\ref{eq:Mnu_pred}).

The light neutrino mass matrix $M_\nu^{}$ can be diagonalized as $M_\nu^{} = V m_\nu^d V^T$, where $m_\nu^d = \mathrm{diag}\{m_1^{}, m_2^{}, m_3^{} \}$ is the diagonalized neutrino mass matrix. In the standard PDG \cite{Patrignani:2016xqp} parameterization, the unitary matrix $V$ can be decomposed as 
\begin{align}\label{eq:pmns}
V =  P_l \left(
\begin{matrix}
c^{}_{12} c^{}_{13} & s^{}_{12} c^{}_{13} & s^{}_{13} e^{-{ i} \delta} \cr 
 -s^{}_{12} c^{}_{23} - c^{}_{12} s^{}_{13} s^{}_{23} e^{{ i} \delta} & c^{}_{12} c^{}_{23} -
s^{}_{12} s^{}_{13} s^{}_{23} e^{{ i} \delta} & c^{}_{13}
s^{}_{23} \cr 
 s^{}_{12} s^{}_{23} - c^{}_{12} s^{}_{13} c^{}_{23}
e^{{ i} \delta} & - c^{}_{12} s^{}_{23} - s^{}_{12} s^{}_{13}
c^{}_{23} e^{{ i} \delta} &   c^{}_{13} c^{}_{23} \cr
\end{matrix} \right) P_{\nu}, \;
\end{align}
where $c^{}_{ij} (s^{}_{ij})$ (for $j=12,23,13$) stands for $\cos\theta^{}_{ij} (\sin\theta^{}_{ij})$, $ P_l^{} = \mathrm{diag}\{e^{i \phi_{e}},e^{i \phi_{\mu}},e^{i \phi_{\tau}} \}$ contains three unphysical phases which can be absorbed by the rephasing of charged lepton fields, and finally $ P_{\nu}^{} = \mathrm{diag}\{e^{i \rho},e^{i \sigma},1\}$ is the Majorana phase matrix. 

Given the form of $M_\nu^{}$, there exist six predictions for the mixing angles and phases introduced above, namely, 
\begin{equation}\label{eq:prediction}
\phi_{e} = 90^\circ,~~~ \phi_{\mu} \equiv - \phi_{\tau}=\phi,~~~ \theta_{23} = 45^\circ,~~~ \delta=\pm 90^\circ,~~~ \rho,~\sigma = 0 ~~{\rm or}~~ 90^\circ.
\end{equation}
A detailed derivation of these predictions is given in Appendix~\ref{app:convention}. Note that in the minimal seesaw framework the lightest neutrino is always massless, and consequently one can always remove one of the Majorana phases. Here we take $\rho$ to be absent. 

As the $\mu-\tau$ reflection symmetry does not specify the values of $\theta_{12}^{}$ and $\theta_{13}^{}$, we proceed to express $\theta_{12}^{}$ and $\theta_{13}^{}$ as follows, 
\begin{align}
\tan \theta^{}_{13} & = \mp ~  
\frac{1}{\sqrt{2}} \frac{{\rm Im} \left(C^{\prime}\right)}
{{\rm Im} \left(B^{\prime}\right) } \; ,  \nonumber \\
%
\tan 2\theta^{}_{12} & = 
\begin{cases}
 \dfrac{2\sqrt{2} \cos 2\theta_{13} {\rm Im} \left(B^{\prime}\right)}{c_{13}\left[({\rm Re }(C^{\prime}) - D) \cos 2\theta_{13} -({\rm Re }(C^{\prime}) + D)s^{2}_{13} + A  c^{2}_{13}\right] } ~; ~~{\rm for }  ~~ {\rm NH }  \;  \\
\dfrac{2\sqrt{2} {\rm Im} \left(B^{\prime}\right)  s^{2}_{13} }{c_{13}\left[{\rm Re } (C^{\prime}) (1 + s^{2}_{13}) + D c^{2}_{13}\right] } ~; ~~{\rm for }  ~~ {\rm IH }
\end{cases}
\end{align}
where $C^{\prime}=C e^{-2i\phi}, B^{\prime}=B e^{-i \phi} $ and the ``$\mp$" sign in $\tan\theta_{13}^{}$ is for $\delta = \pm 90^\circ$. Lastly, for the light neutrino masses we have 
\begin{eqnarray}
m_1 = 0, \quad m_2^{} e^{2i\sigma} = \dfrac{2\sqrt{2} {\rm Im}(B^{\prime}) }{c_{13}\sin 2 \theta_{12}}, \quad m_3 = A + 2D +  \dfrac{2\sqrt{2} {\rm Im}(B^{\prime}) }{c_{13}\sin 2 \theta_{12}}  \; ,
\end{eqnarray}
for the NH case, while in IH the neutrino masses turn out to be
\begin{eqnarray}
m_1 =  D - \dfrac{A}{2} -  \dfrac{ \sqrt{2} {\rm Im}(B^{\prime}) }{c_{13}\sin 2 \theta_{12}}, \quad m_2^{} e^{2i\sigma} = - D - \dfrac{A}{2} + \dfrac{\sqrt{2} {\rm Im}(B^{\prime}) }{c_{13}\sin 2 \theta_{12}}, \quad m_3 = 0 \; .
\end{eqnarray}
We notice that in both NH and IH there exists a relation among neutrino masses, i.e.,
\begin{equation}
-m_1 - m_2 e^{2i\sigma} + m_3 = A + 2D \; .
\end{equation}

Having introduced the $\mu-\tau$ reflection symmetry in the minimal seesaw formalism, in the subsequent sections we study the  breaking of such a symmetry and its impact on neutrino oscillation parameters at low energies.

\section{Breaking due to Renormalization Group Running} \label{sec:rge}

We start with the investigation on the breaking of $\mu-\tau$ reflection symmetry due to the RG running. As one possible ultraviolet extension of the SM, the minimal supersymmetric standard model (MSSM) is taken to be our theoretical framework at high energies.\footnote{{We here do not consider the RG running in the SM, as in general the running effects in the SM are weaker than those in MSSM~\cite{Antusch:2003kp}.}} 
Within MSSM, the neutrino Yukawa coupling in Eq.~(\ref{eq:lag}) needs to be modified to $\overline{\nu_L^{}} Y_\nu^{} N_R^{} H_u^{}$, where $H_u^{}$ is the Higgs field that also couples to the up-quark sector. When $H_u^{}$ picking up the vacuum expectation value, i.e., $\langle H_u^{} \rangle = v_u^{} = v \sin\beta$, the neutrino Dirac mass matrix $M_D^{}$ becomes as $M_D^{} = v \sin\beta Y_\nu^{}$. Moreover, we take the scale of grand unified theories (GUTs) ($\Lambda_\mathrm{GUT}^{}$) as the high energy boundary scale, at which the $\mu-\tau$ reflection symmetry is viewed to be exact in $Y_\nu^{}$ (or $M_D^{}$) and $M_R^{}$. 

The RG running towards low energies can then be divided into three stages. The first stage of running starts from the GUT scale and ends at the mass threshold of the heavier right-handed neutrino $N_2^{}$, schematically, $\Lambda_\mathrm{GUT}^{} \rightarrow M_2^{}$. The one-loop RG equations of relevant Yukawa and mass matrices are given by~\cite{Antusch:2003kp, Antusch:2005gp,Mei:2005qp}
\begin{eqnarray} \label{eq:RG_Yl}
 \frac{\mathrm{d} Y_l^{}}{\mathrm{d} t} &=& \left( \alpha_l^{} + 3 Y_l^{} Y_l^\dagger + Y_\nu^{} Y_\nu^\dagger \right) Y_l^{} \;, \\
\label{eq:RG_Ynu}
 \frac{\mathrm{d} Y_\nu^{}}{\mathrm{d} t} &=& \left( \alpha_\nu^{} + Y_l^{} Y_l^\dagger + 3 Y_\nu^{} Y_\nu^\dagger \right) Y_\nu^{} \;, \\
\label{eq:RG_MR}
\frac{\mathrm{d} M_R^{}}{\mathrm{d} t} & = & 2 \left[ M_R^{} \left( Y_\nu^\dagger Y_\nu^{} \right) + \left( Y_\nu^\dagger Y_\nu^{} \right)^T M_R^{} \right] \; ,
\end{eqnarray}
where $t=\ln (\mu / \Lambda_{\mathrm{GUT}}^{}) / (16\pi^2)$ with $\mu$ being the renormalization scale, and the flavor-independent parameters $\alpha_l^{}$ and $\alpha_\nu^{}$ are defined as
\begin{eqnarray}
\alpha_l^{} &\equiv & \mathrm{Tr} ( 3 Y_d^{} Y_d^\dagger + Y_l^{} Y_l^\dagger  ) - \left( \frac{9}{5}g_1^2 + 3 g_2^2 \right) \; , \\
\alpha_\nu^{} &\equiv & \mathrm{Tr} (3 Y_u^{} Y_u^\dagger + Y_\nu^{} Y_\nu^\dagger ) - \left( \frac{3}{5}g_1^2 + 3 g_2^2 \right) \; .
\end{eqnarray}
In the above $Y_u^{}$, $Y_d^{}$ and $Y_l^{}$ are the up-quark, down-quark and charged-lepton Yukawa matrices, respectively.  We note that unlike the RG running below the seesaw threshold, the charged-lepton Yukawa matrix $Y_l^{}$ now could be non-diagonal due to the term of $Y_\nu^{} Y_\nu^\dagger$, even if it were diagonal at the high energy boundary. As a result, when extracting the lepton mixing parameters above the seesaw threshold, one also needs to take into account the corrections from $Y_l^{}$. The light neutrino mass matrix at this stage of running is given by $M_\nu^{(2)} = - v^2 \sin^2\beta~Y_\nu^{} M_R^{-1} Y_\nu^T$, and the RG running of $M_\nu^{(2)}$ is found to be~\cite{Antusch:2003kp, Antusch:2005gp,Mei:2005qp}
\begin{eqnarray} \label{eq:RG_Mnu}
\frac{\mathrm{d} M_\nu^{(2)}}{\mathrm{d} t} = 2 \alpha_\nu^{} M_\nu^{(2)} + \left( Y_l^{} Y_l^\dagger + Y_\nu^{} Y_\nu^\dagger \right) M_\nu^{(2)} + M_\nu^{(2)} \left( Y_l^{} Y_l^\dagger + Y_\nu^{} Y_\nu^\dagger \right)^T \; . 
\end{eqnarray}
Similar to $Y_l^{}$, the evolution of $M_\nu^{(2)}$ now also involves the contribution from the neutrino Yukawa matrix $Y_\nu^{}$. 

When the renormalization scale is below the mass threshold of $N_2^{}$, the second stage of running starts and it ends at the mass threshold of $N_1^{}$, i.e., $M_2^{} \rightarrow M_1^{}$. At the matching scale $\mu = M_2^{}$, the light neutrino mass matrix is given by
\begin{eqnarray}
M_\nu^{(1)} = - v^2 \sin^2\beta~\left( \widehat{Y}_\nu^{} \widehat{M}_R^{-1} \widehat{Y}_\nu^T + \widetilde{Y}_\nu^{} M_2^{-1} \widetilde{Y}_\nu^T  \right) \; ,
\end{eqnarray}
where $\widehat{Y}_\nu^{}$ and $\widehat{M}_R^{}$ are the $Y_\nu^{}$ and $M_R^{}$ with the entries corresponding to $N_2^{}$ removed, while $\widetilde{Y}_\nu^{}$ is the column corresponding to $N_2^{}$ in $Y_\nu^{}$. Specifically, $\widehat{Y}_\nu^{}(\mu = M_2^{})$ and $\widetilde{Y}_\nu^{}(\mu = M_2^{})$ are the first and second columns of $Y_\nu^{} (\mu = M_2^{})$, respectively, and $\widehat{M}_R^{}(\mu = M_2^{})$ is the (11) entry of $M_R^{} (\mu = M_2^{})$. Below $\mu = M_2^{}$, the one-loop running of $Y_l^{}$, $Y_\nu^{}$, $M_R^{}$ and $M_\nu^{}$ is still formally dictated by Eqs.~(\ref{eq:RG_Yl},\ref{eq:RG_Ynu},\ref{eq:RG_MR},\ref{eq:RG_Mnu}), except that we replace $Y_\nu^{}$ and $M_R^{}$ by $\widehat{Y}_\nu^{}$ and $\widehat{M}_R^{}$. 

The final stage of RG running starts from the mass threshold of $N_1^{}$ and stops at a chosen low energy scale. Here we take the low energy scale to be the electroweak scale $\Lambda_{\mathrm{EW}}^{}$. This stage of RG running is below the seesaw threshold, and its impact on the lepton mixing parameters has been extensively discussed in the literature, e.g., Refs.~\cite{Antusch:2003kp, Antusch:2005gp,Mei:2005qp,Ohlsson:2013xva}. In particular, the breaking of $\mu-\tau$ reflection symmetry due to this stage of RG running is investigated in Refs.~\cite{Zhou:2014sya,Zhao:2017yvw}. To save space, we then would not elaborate more on this stage of running. 

With the above RG equations, in principle one can investigate the breaking of $\mu-\tau$ reflection symmetry due to the RG running analytically, as was done in Ref.~\cite{King:2016yef} for the littlest seesaw scenario. However, in the current setup all entries of $Y_\nu^{}$ are non-zero, so that an analytical study turns out to be formidable. We then choose to study this issue numerically. In the numerical study, we set $\tan\beta = 30$, and the high and low energy boundary scales are taken to be $\Lambda_{\mathrm{GUT}}^{} = 2 \times 10^{16}~\mathrm{GeV}$ and $\Lambda_{\mathrm{EW}}^{} = 1~\mathrm{TeV}$, respectively.  We also study the cases considering $\tan\beta < 30$, and as the modifications on mixing parameters are quite small we do not include those results here.
The gauge couplings and various Yukawa couplings at $\Lambda_{\mathrm{GUT}}^{}$ are taken to be the default values in the numerical RG running package $\texttt{REAP}$~\cite{Antusch:2005gp}, although $M_D^{}$ (or $Y_\nu^{}$) and $M_R^{}$ are set to be the forms given in Eq.~(\ref{eq:md}) and Eq.~(\ref{eq:mr}), respectively. Our numerical strategy is to scan all the parameters in $Y_\nu^{}$ and $M_R^{}$ at high energies, and seek the allowed parameter space that yields lepton mixing parameters compatible with current experimental data at low energies. The varied ranges of parameters at the high energy boundary are as follows,
\begin{eqnarray}
|b|, |c|, |d| \in [0, 1]~v, \quad (M_{R}^{})_{22,23}^{} \in [10^{12}, 10^{15}]~\mathrm{GeV}, \quad \phi_{b,c,d,M} \in [0, 2\pi) \; .
\end{eqnarray}
To guide the parameter scan, we employ the nested sampling package $\texttt{Multinest}$ \cite{Feroz:2007kg,Feroz:2008xx,Feroz:2013hea}, with a $\chi^2$ function built based on the latest global fit results~\cite{deSalas:2017kay}. Details about the $\chi^2$ function can be found in Appendix~\ref{app:chi_sq}. 

\begin{figure}
\centering
\includegraphics[scale=0.4]{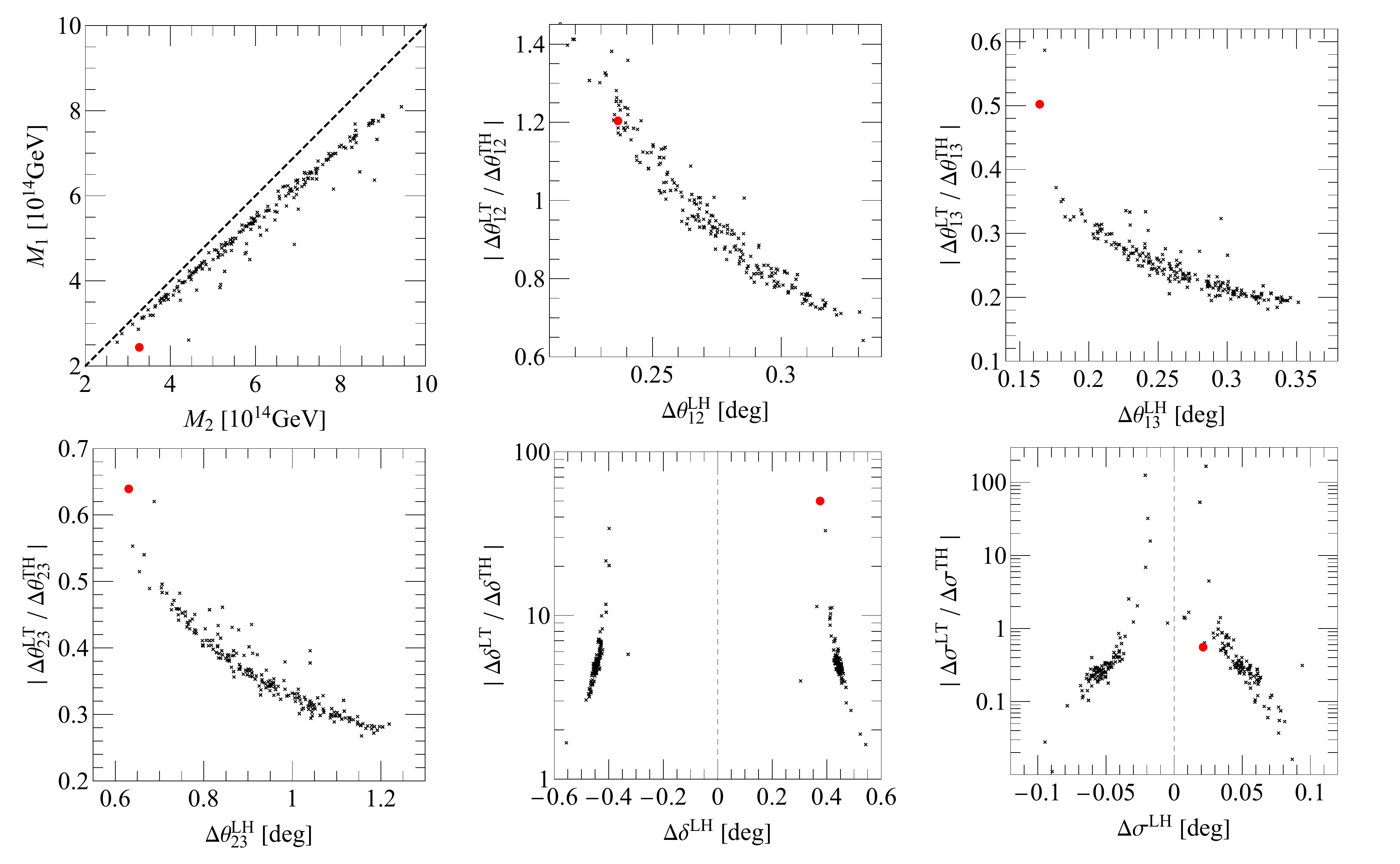}
\caption{\footnotesize Predictions of RGE running in NH under the framework of MSSM with $\tan\beta = 30$. }
\label{fg:RGE_NH_comb}
\end{figure}

In Fig.~\ref{fg:RGE_NH_comb} we show the numerical result for the NH case. The black scatter points have $\chi^2 < 30$, and the best-fit (BF) point that has the minimal value of $\chi^2$, denoted as $\chi^2_\mathrm{min}$, is shown in red. In this NH case, we obtain $\chi^2_\mathrm{min} = 19.34$ for the BF point. In the $M_2^{}$ vs. $M_1^{}$ plot of Fig.~\ref{fg:RGE_NH_comb}, we display the spread of two mass thresholds $M_{1,2}^{}$ for the right-handed neutrinos. It can be seen that $M_1^{}$ and $M_2^{}$ are quite to close  each other and both of the order of $10^{14}~\mathrm{GeV}$. One possible explanation for such closeness of $M_1^{}$ and $M_2^{}$ is that the entries in two columns of $Y_\nu^{}$ are related by the $\mu-\tau$ reflection symmetry, particularly due to the symmetry transformation on $N_R^{}$. Therefore, no large hierarchy exists between the two columns of $Y_\nu^{}$, and then in order to yield mild hierarchy in the light neutrino mass matrix, the entries in $M_R^{}$ also tend to be close to each other, resulting in similar values of $M_1^{}$ and $M_2^{}$. Because of the closeness of $M_1^{}$ and $M_2^{}$, the second stage of RG running between two mass thresholds turns out to be insignificant, and thus we focus on the first and third stages of running in the following.

For the convenience of quantifying the RG running effects, we introduce the quantities of $\Delta x^{\mathrm{LH}}$, $\Delta x^{\mathrm{LT}}$ and $\Delta x^{\mathrm{TH}}$ in the rest plots of Fig.~\ref{fg:RGE_NH_comb}. Here $x$ stands for the lepton mixing angles and phases, and we define $\Delta x^{\mathrm{LH}}$ as the difference of the mixing parameter $x$ at the low and high energy scales, i.e., $\Delta x^{\mathrm{LH}} \equiv x(\Lambda_\mathrm{EW}^{}) - x( \Lambda_\mathrm{GUT}^{})$. Similarly, we have $\Delta x^{\mathrm{LT}} \equiv x(\Lambda_\mathrm{EW}^{}) - x(M_2^{})$ and $\Delta x^{\mathrm{TH}} \equiv x(M_2^{}) - x(\Lambda_\mathrm{GUT}^{})$. To compare the RG running effects between the first and third stages of running, in the y-axis of these plots we show the absolute values of the ratios of $\Delta x^{\mathrm{LT}} / \Delta x^{\mathrm{TH}}$. By inspecting Fig.~\ref{fg:RGE_NH_comb}, we then observe:
\begin{itemize}
\item  In this NH case all $\Delta x^\mathrm{LH}$'s (shown as the x-axis) are rather small, indicating the mixing angles and phases receive small deviations from the RG running. The small deviations at the third stage of RG running are expected, as it is known that in NH the RG running of mixing angles and phases are insignificant below the seesaw threshold~\cite{Antusch:2003kp, Antusch:2005gp,Mei:2005qp}. For the first stage of running, the corrections to $M_\nu^{(2)}$ are at the order of $ \ln (\Lambda_\mathrm{GUT} / M_2^{}) /(16\pi^2) \sim 0.03$, assuming $Y_\nu^{}$ to be of $\mathcal{O}(1)$. Therefore, in the NH case the contributions from the first stage of running are also small. 

\item Regarding the relative contributions between the first and third stages of running, we notice that for the Dirac phase $\delta$, the third stage of running tends to yield larger deviations than the first stage, as $|\Delta x^{\mathrm{LT}} / \Delta x^\mathrm{TH} | > 1$ for most of scatter points. However, for the three mixing angles and the Majorana phase $\sigma$, the first stage of running turns out to be more important than, or as important as, the first stage. It then demonstrates that in this seesaw embedded setup, the RG running effects above the seesaw threshold can be comparable with that below the seesaw threshold.

\item As for the breaking of $\mu-\tau$ reflection symmetry, $\theta_{23}^{}$ at low energies tends to be always larger than $45^\circ$, while $\delta$ and $\sigma$ can receive either positive or negative deviations from RG running. The correlation between the positive deviation of $\theta_{23}^{}$ and NH is in agreement with the previous RG running studies below the seesaw threshold
~\cite{Luo:2014upa,Zhang:2016png}. 

\end{itemize}

\begin{figure}
\centering
\includegraphics[scale=0.47]{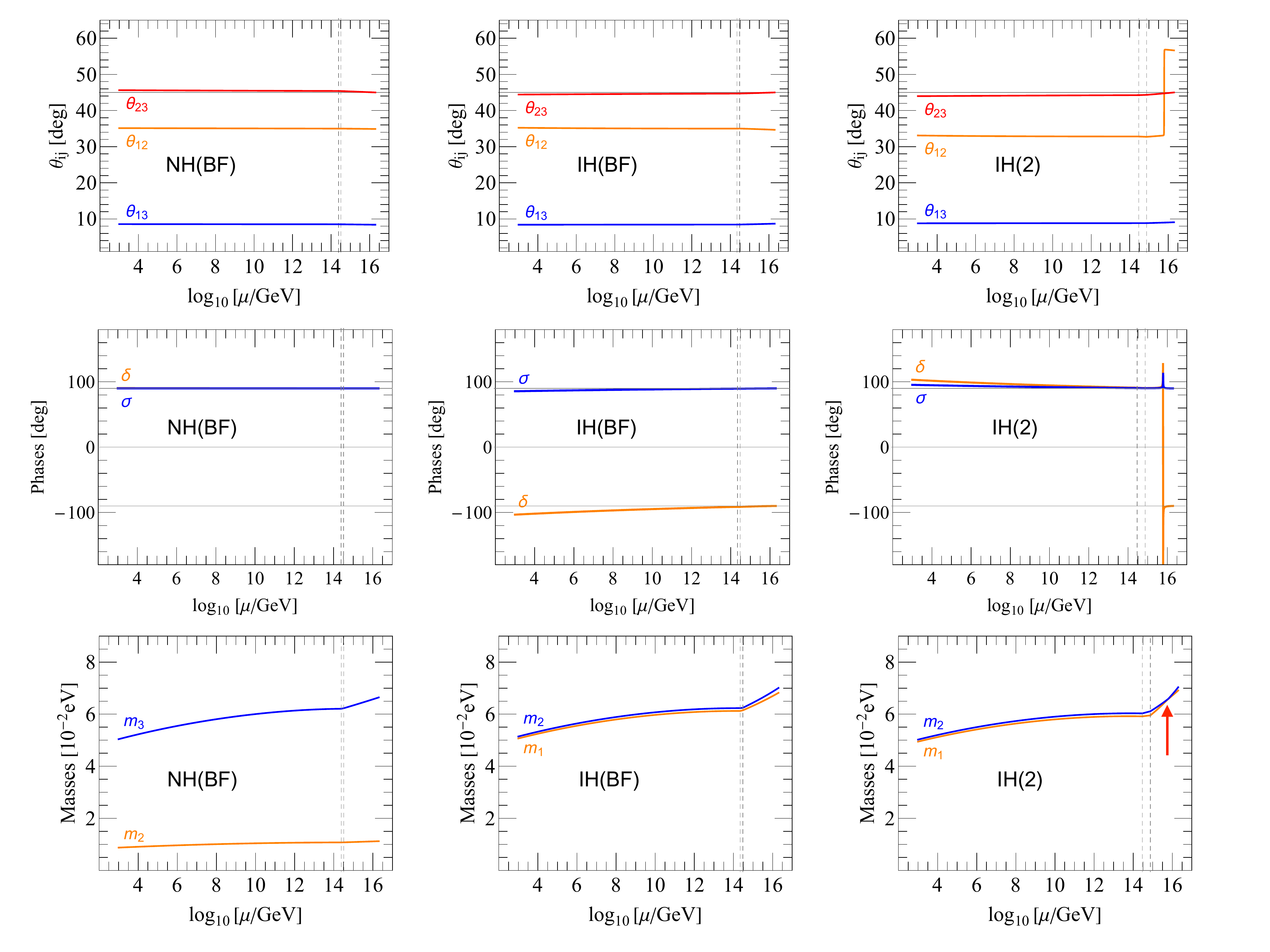}
\caption{\footnotesize The detailed RGE running of mixing parameters for the best-fit scenarios in NH (left) and IH (middle), and an alternative scenario in IH (right). Dashed vertical lines denote the locations of the mass thresholds of $N_1^{}$ and $N_2^{}$, and the red arrow indicates the scale where the two light neutrino masses become degenerate. Note that the Majorana phase $\sigma$ is taken to be within $[0, \pi)$ by convention.}
\label{fg:RGE_BF}
\end{figure}

To have better feeling of the RG running in this NH case, in the left three plots of Fig.~\ref{fg:RGE_BF} we show the detailed RG running of the mixing angles, phases and neutrino masses for the BF point. One can easily see that the two mass thresholds, indicated by the dashed vertical lines, are quite close to each other, and the running of mixing angle and phases are indeed not appreciable. However, significant running is observed for the neutrino masses, and because there exist contributions from $Y_\nu^{}$ in $\alpha_\nu^{}$ during the first stage of running, RG running at the first stage is more dramatic than the third stage.

\begin{figure}
\centering
\includegraphics[scale=0.4]{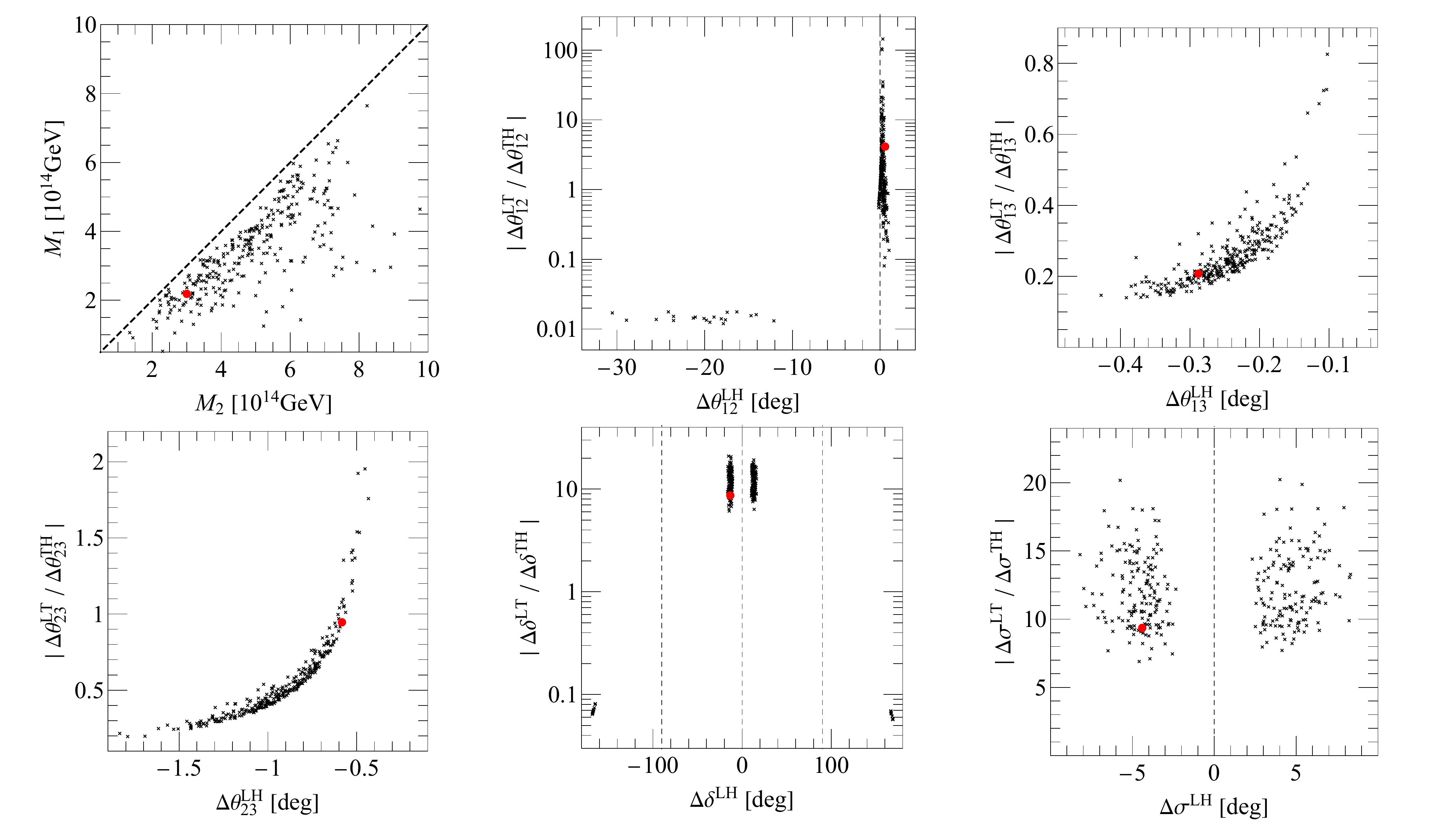}
\caption{\footnotesize Predictions of RGE running in IH under the framework of MSSM with $\tan\beta = 30$. }
\label{fg:RGE_IH_comb}
\end{figure}

We next turn to the IH case, and the corresponding numerical results are shown in Fig.~\ref{fg:RGE_IH_comb}. We find that in IH it becomes harder to search for the scatter points that have small values of $\chi^2$. Thus, here we show the scatter points that satisfy $\chi^2 < 80$, and the BF point has $\chi^2_\mathrm{min} = 41.75$. The detailed RG running of mixing parameters for the BF point is shown in the middle three plots of Fig.~\ref{fg:RGE_BF}. By inspecting plots in Fig.~\ref{fg:RGE_IH_comb}, we first observe that $M_1^{}$ and $M_2^{}$ are also close to each other, as in the NH case. The running of the three mixing angles is again quite mild, except that in $\theta_{12}^{}$ there exists a branch of scatter points that can have deviations as large as $30^\circ$. To have better understanding of such large deviations in $\theta_{12}^{}$, in the right three plots of Fig.~\ref{fg:RGE_BF} we present the detailed RG running for the scatter point that has the least value of $\chi^2$ ($\chi^2_\mathrm{min} = 51.3$) in that branch of scatter points. One then identifies a sharp decline in $\theta_{12}^{}$ during the first stage of running. Such a decline can be traced to the potential crossing of neutrino masses when $\mu \sim 5 \times 10^{15}~\mathrm{GeV}$ (red arrow). As interchanging the order of eigenvalues would lead to a $90^\circ$ rotation in the mixing angle, it also explains why the sum of the values of $\theta_{12}^{}$ before and after the decline is around $90^\circ$. From the running plot of $\delta$, such interchange of eigenvalues seems to induce a $180^\circ$ change in $\delta$ as well. 

As for the Majorana phase $\sigma$, from Fig.~\ref{fg:RGE_IH_comb} we notice that the running of $\sigma$ is also quite mild in this IH case. Moreover, we find that the obtained values of $\sigma$'s for the scatter points shown here are all around $90^\circ$. Although $\sigma = 0$ is also predicted by the $\mu-\tau$ reflection symmetry, having $\sigma = 0$ at the high energy boundary would lead to dramatic running in $\theta_{12}^{}$.  This can be seen from the following RG equation for $\theta_{12}^{}$ below the seesaw threshold~\cite{Antusch:2003kp, Antusch:2005gp,Mei:2005qp,Dighe:2007ksa,Ohlsson:2013xva},
\begin{eqnarray}
\frac{\mathrm{d} \theta_{12}^{}}{\mathrm{d} t} \propto \sin 2\theta_{12}^{} \sin^2\theta_{23} \frac{|m_1^{} + m_2^{} e^{2 i \sigma}|^2}{\Delta m_{21}^2} + \mathcal{O}(\theta_{13}^{}) \; .
\end{eqnarray}
If $\sigma \sim 0$, there then exists an enhancement factor of $(m_1^{} + m_2^{})/ \Delta m_{21}^2$. Such dramatic running of $\theta_{12}^{}$ would hinder the sampling program to map out the allowed parameter space corresponding to $\sigma \sim 0$ at high energies. In contrast, if $\sigma \sim 90^\circ$ and because $m_1^{} \sim m_2^{}$ in IH, the combination of $|m_1^{} + m_2^{} e^{2i \sigma}|$ becomes vanishingly small. This also explains why even in IH  the running of $\theta_{12}^{}$ is still insignificant.

Lastly, we point out that in this IH case $\theta_{23}^{}$ tends to be smaller than $45^\circ$ at low energy. As in the NH case, this observed correlation between the negative deviation of $\theta_{23}^{}$ and IH is in agreement with the previous RG running studies below the seesaw threshold~\cite{Luo:2014upa,Zhang:2016png}.  

In Appendix~\ref{app:rge_table} we present the numerical values of the neutrino Yukawa matrices and the Majorana mass matrices for the right-handed neutrinos at $\Lambda_\mathrm{GUT}^{}$ for the three scenarios shown in Fig.~\ref{fg:RGE_BF}. The lepton mixing parameters at various energy scales are also shown. 

\section{Breaking $\mu-\tau$ reflection symmetry in $M_D^{}$ and $M_R^{}$}\label{ses:symm_break}

From the previous RG running study we notice that in both NH and IH the breaking effects due to the RG running are quite mild. For instance, the deviations in $\theta_{23}^{}$ are only around one degree. Although such small deviations are in compatible with current experimental data, it may become necessary to consider large deviations when more accurate data will be included. In this section, we set out to discuss the breaking of $\mu-\tau$ reflection symmetry in the low energy neutrino mass matrix by introducing explicit breaking terms in the neutrino Dirac mass matrix $M_D^{}$ and the Majorana mass matrix $M_R^{}$ for the right-handed neutrinos.  As the RG running effects are found to be mild, for simplicity we choose to ignore them in the following discussion. 

\subsection{Breaking $\mu-\tau$ reflection symmetry in $M_D^{}$}

We start with assigning an explicit breaking term in the $(12)$ position of $M_D^{}$, so that the neutrino Dirac mass matrix $M_D^\prime$ after breaking is given by
\begin{eqnarray}\label{eq:md_br}
\mathbf{S1:}\quad M_D^\prime = \left( \begin{matrix}b & b^* (1 + \epsilon) \cr
  c & d \cr d^* & c^* 
\end{matrix} \right) \; ,
\end{eqnarray}
where $\epsilon$ is a small breaking parameter, taken to be real for simplicity. We name this breaking scenario as $\mathbf{S1}$. The above  $M_D^\prime$ leads to a new mass matrix $M_\nu^\prime$ for the light neutrinos, and the difference between $M_\nu^\prime$ and $M_\nu^{}$ is given by
\begin{eqnarray}\label{eq:refle_br_12}
\Delta M_{\nu}^{} \equiv M_\nu^\prime - M_\nu^{} =    \epsilon \mathcal{B}_{12}^{} \begin{pmatrix} 
2 \widehat{A}_1^{} & \widehat{A}_2^{} & \widehat{A}_3^{} \\
 \widehat{A}_2^{} & 0 & 0 \\
 \widehat{A}_3^{} & 0 & 0 \\
\end{pmatrix} + \mathcal{O}(\epsilon^2_{})
\end{eqnarray}
where $\mathcal{B}_{12}^{} = b^*_{} /{\rm det}(M_R)$, and $\widehat{A}_i^{}$'s are defined as
\begin{eqnarray} \label{eq:Aheads}
\widehat{A}_1^{} = b^{*}_{} m_{22}^{} - b m_{23}^{} \; ,  \quad
\widehat{A}_2^{} = d m_{22}^{} - c m_{23}^{} \; ,\quad
\widehat{A}_3^{} = c^{*} m_{22}^{} - d^* m_{23}^{} \; .
\end{eqnarray}

To evaluate the impact of the above breaking on the neutrino masses and lepton mixing angles, we diagonalize $M_\nu^\prime$ with the mixing matrix $V^\prime$, which coincides with the mixing matrix $V$ when $\epsilon = 0$. For simplicity, we consider the scenario in which all entries of $\Delta M_\nu^{}$ are real, and expand the deviations of neutrino masses and lepton mixing angles in terms of small parameters $\epsilon$, $\theta_{13}^{}$ and $\zeta = m_2^{}/m_3^{}$ ($\xi = \Delta m_{21}^2 / m_2^2$) for NH (IH). In the top block of Table~\ref{tab:break_YD} we show the leading order results for the deviations of neutrino masses $\Delta m_i^{} = m_i^\prime - m_i^{}$ (for $i= 1, 2, 3$) and the deviations of lepton mixing angles $\Delta \theta_{ij}^{} = \theta_{ij}^\prime - \theta_{ij}^{}$  (for $ij= 12, 13, 23$), where $m_i^\prime$'s and $\theta_{ij}^\prime$'s are the neutrino masses and mixing angles after breaking. It can be seen that in both NH and IH cases, because of the factor $\theta_{13}^{}$ in $\Delta \theta_{23}^{}$ and the factor $1/\zeta \sim 5$ or $1/\xi \sim 30$ in $\Delta \theta_{12}^{}$,  we have $|\Delta \theta_{23}^{}| < |\Delta \theta_{13}^{}| < |\Delta \theta_{12}^{}|$ in general, barring the cases that accidental cancellations exist among $\widehat{A}_i^{}$'s. The suppression of $\Delta \theta_{23}^{}$ also indicates that even with the breaking term in the (12) position of $M_D^{}$, the predicted $\theta_{23}^{}$ after breaking is still quite close to $45^\circ$. 

\begin{table}
\centering
\scriptsize
\begin{tabular}{c | c  | c | c}
\hline
\hline
&  & NH & IH \\
\hline
\multirow{3	}{*}{{\normalsize{$\mathbf{S1:}$}}~$M_D^{} = \begin{pmatrix}b & b^* (1+\epsilon) \cr
  c & d \cr d^* & c^* 
\end{pmatrix}   $} & $\Delta m_1^{}$ & 0 & $\epsilon \mathcal{B}_{12}^{} \left[2\widehat{A}_1^{} c_{12}^{2} - \sqrt{2} \widehat{A}_{(23)}^{} s_{12}^{} c_{12}^{} \sin\phi \right ]$\\
& $\Delta m_2^{}$ & $\epsilon \mathcal{B}_{12}^{} \left[2\widehat A_1^{} s_{12}^{2} + \sqrt{2} \widehat{A}_{(23)}^{} s_{12}^{} c_{12}^{} \sin\phi \right ]$ & $\epsilon \mathcal{B}_{12}^{} \left[ 2\widehat A_1^{} s_{12}^{2} + \sqrt{2} \widehat{A}_{(23)}^{} s_{12}^{} c_{12}^{} \sin\phi \right ]$ \\
& $\Delta m_3^{}$ & $\sqrt{2}  \epsilon \theta_{13}^{} \mathcal{B}_{12}^{}  \widehat{A}_{(23)}^{}  \cos \phi $ & 0 \\
\multirow{3	}{*}{$\Delta M_\nu^{} \simeq \epsilon \mathcal{B}_{12}^{} \begin{pmatrix} 2 \widehat A_1^{} & \widehat A_2^{} & \widehat A_3^{} \cr
  \widehat A_2 & 0 & 0 \cr
  \widehat A_3 & 0 & 0 \cr
\end{pmatrix} $} & $\Delta \theta_{12}^{}$ & $\frac{\epsilon \mathcal{B}_{12}^{}}{2 m_3^{} \zeta} \left[ \widehat{A}_1^{} \sin 2 \theta_{12}^{} + \sqrt{2}  \widehat{A}_{(23)}^{} \cos 2\theta_{12}^{}  \sin\phi  \right]$  & $\frac{\epsilon \mathcal{B}_{12}^{}}{ m_2^{} \xi} \left[2 \widehat{A}_1 \sin 2\theta_{12}^{} + \sqrt{2}  \widehat{A}_{(23)}^{} \cos 2\theta_{12}^{}  \sin\phi  \right]$ \\
& $\Delta \theta_{13}^{}$ & $ \frac{\epsilon \mathcal{B}_{12}^{}}{\sqrt{2} m_3^{}}  \widehat{A}_{(23)}^{} \cos\phi$ &   $- \frac{\epsilon \mathcal{B}_{12}^{}}{\sqrt{2} m_2^{}}  \widehat{A}_{(23)}^{} \cos\phi$\\
& $\Delta \theta_{23}^{}$ &  $ \frac{\epsilon \mathcal{B}_{12}^{}}{\sqrt{2} m_3^{}}   \theta_{13}^{} \widehat{A}_{[23]}^{} \cos\phi$ & $ \frac{\epsilon \mathcal{B}_{12}^{} }{\sqrt{2} m_2^{}}   \theta_{13}^{}  \widehat{A}_{[23]}^{} \cos \phi$ \\
\hline
\hline
\multirow{3	}{*}{{\normalsize{$\mathbf{S2:}$}}~$M_D^{} = \begin{pmatrix}b & b^*  \cr
  c & d (1+\epsilon) \cr d^* & c^* 
\end{pmatrix}   $}  & $\Delta m_1^{}$ & 0 & $-\epsilon \mathcal{B}_{22}^{} \left[ \widehat{A}_{[23\phi]}^{} s^2_{12}  + \sqrt{2} \widehat{A}_1^{} s_{12}^{} c_{12}^{}  \sin \phi \right ]$\\
& $\Delta m_2^{}$ & $-\epsilon \mathcal{B}_{22}^{} \left[  \widehat{A}_{[23\phi]}^{} c^2_{12}  - \sqrt{2} \widehat{A}_1^{} s_{12}^{} c_{12}^{}  \sin \phi \right ]$ & $-\epsilon \mathcal{B}_{22}^{} \left[  \widehat{A}_{[23\phi]}^{} s^2_{12}  - \sqrt{2} \widehat{A}_1^{} s_{12}^{} c_{12}^{}  \sin \phi \right ]$ \\
& $\Delta m_3^{}$ & $\epsilon  \mathcal{B}_{22}^{} \widehat{A}_{(23\phi)}^{}$ & 0 \\
\multirow{3	}{*}{$\Delta M_\nu^{} \simeq \epsilon \mathcal{B}_{22}^{} \begin{pmatrix} 
0 & \widehat{A}_1^{} & 0 \\
\widehat{A}_1^{} & 2 \widehat{A}_3^{} & \widehat{A}_2^{} \\
0 & \widehat{A}_2^{} & 0
\end{pmatrix} $}  & $\Delta \theta_{12}^{}$ & $\frac{\epsilon \mathcal{B}_{22}^{}}{2 m_3^{} \zeta} \left[  \widehat{A}_{[23\phi]}^{} \sin 2\theta_{12}^{} + \sqrt{2} \widehat{A}_1^{} \cos 2\theta_{12}^{}  \sin\phi  \right]$  &  $\frac{\epsilon \mathcal{B}_{22}^{}}{m_2^{} \xi} \left[ \widehat{A}_{[23\phi]}^{} \sin 2\theta_{12}^{} + \sqrt{2} \widehat{A}_1^{} \sin\phi \cos 2\theta_{12}^{} \right]$ \\
& $\Delta \theta_{13}^{}$ & $ \frac{\epsilon \mathcal{B}_{22}^{}}{\sqrt{2} m_3^{}} \widehat{A}_1^{} \cos\phi$ & $ -\frac{\epsilon \mathcal{B}_{22}^{}}{\sqrt{2} m_2^{}} \widehat{A}_1^{} \cos\phi $ \\
& $\Delta \theta_{23}^{}$ &  $- \frac{\epsilon \mathcal{B}_{22}^{}}{ m_3^{}}\widehat{A}_3^{}\cos 2\phi$ & $ \frac{\epsilon \mathcal{B}_{22}^{}}{m_2^{}} \widehat{A}_3^{}\cos 2\phi$ \\
\hline
\hline
\multirow{3	}{*}{{\normalsize{$\mathbf{S3:}$}}~$M_D^{} = \begin{pmatrix}b & b^*  \cr
  c & d \cr d^* & c^*  (1+\epsilon)
\end{pmatrix}   $} & $\Delta m_1^{}$ & 0 & $-\epsilon \mathcal{B}_{32}^{} \left[ \widehat{A}_{[23\phi]} s^2_{12}  + \sqrt{2} \widehat{A}_1^{} s_{12}^{} c_{12}^{}  \sin \phi \right ]$\\
& $\Delta m_2^{}$ & $-\epsilon \mathcal{B}_{32}^{} \left[ \widehat{A}_{[23\phi]}  c^2_{12}  - \sqrt{2} \widehat{A}_1^{} s_{12}^{} c_{12}^{}  \sin \phi \right ]$ & $-\epsilon \mathcal{B}_{32}^{} \left[  \widehat{A}_{[23\phi]}  c^2_{12}  - \sqrt{2} \widehat{A}_1^{} s_{12}^{} c_{12}^{}  \sin \phi \right ]$ \\
& $\Delta m_3^{}$ & $\epsilon \mathcal{B}_{32}^{}  \widehat{A}_{(23\phi)} $ & 0 \\
\multirow{3	}{*}{$\Delta M_\nu^{} \simeq \epsilon \mathcal{B}_{32}^{} \begin{pmatrix} 
0 & 0 & \widehat{A}_1^{} \\
0 & 0 & \widehat{A}_2^{} \\
\widehat{A}_1^{} & \widehat{A}_2^{} & 2 \widehat{A}_3^{} 
\end{pmatrix} $}  & $\Delta \theta_{12}^{}$ & $ \frac{\epsilon \mathcal{B}_{32}^{}}{2 m_3^{} \zeta} \left[ \widehat{A}_{[23\phi]}  \sin 2\theta_{12}^{} + \sqrt{2} \widehat{A}_1^{} \sin\phi \cos 2\theta_{12}^{} \right]$  & $\frac{\epsilon \mathcal{B}_{32}^{}}{m_2^{} \xi} \left[  \widehat{A}_{[23\phi]}  \sin 2\theta_{12}^{} + \sqrt{2} \widehat{A}_1^{} \sin\phi \cos 2\theta_{12}^{} \right] $\\
& $\Delta \theta_{13}^{}$ & $ \frac{\epsilon \mathcal{B}_{32}^{} }{\sqrt{2} m_3^{}} \widehat{A}_1^{} \cos\phi$ & $ -\frac{\epsilon \mathcal{B}_{32}^{}}{\sqrt{2} m_2^{}} \widehat{A}_1^{} \cos\phi $ \\
& $\Delta \theta_{23}^{}$ & $ -\frac{\epsilon \mathcal{B}_{32}^{}}{m_3^{}} \widehat{A}_3^{} \cos 2 \phi $ & $ -\frac{\epsilon \mathcal{B}_{32}^{}}{m_2^{}} \widehat{A}_3^{} \cos 2 \phi $ \\
\hline
\hline
\end{tabular}
\caption{\footnotesize Corrections to neutrino masses and lepton mixing angles according to the three breaking patterns in $M_D^{}$. For simplicity, we assume that all entries in $\Delta M_\nu^{}$ are real, and only the leading order corrections in terms of $\epsilon$, $\theta_{13}^{}$ and $\zeta = m_2^{}/m_3^{}$ ($\xi = \Delta m_{21}^2 / m_2^2$) for NH (IH) are kept. Short-hand notations of $\widehat{A}_{(23)}^{} \equiv \widehat{A}_2^{} + \widehat{A}_3^{}$, $\widehat{A}_{[23]}^{} \equiv \widehat{A}_2^{} - \widehat{A}_3^{}$, $\widehat{A}_{(23\phi)}^{} \equiv \widehat{A}_3^{} \cos 2\phi + \widehat{A}_2^{}$ and $\widehat{A}_{[23\phi]}^{} \equiv \widehat{A}_3^{} \cos 2\phi - \widehat{A}_2^{}$ are adopted.}
\label{tab:break_YD}
\end{table}

Similarly, one can introduce breaking terms in the other entries of $M_D^{}$. Without loss of generality, in the middle and bottom blocks of Table~\ref{tab:break_YD} we show the results for the other two breaking patterns in $M_D^{}$, namely, assigning breaking terms in the (22) and (32) positions of $M_D^{}$ and resulting in the breaking scenarios of $\mathbf{S2}$ and $\mathbf{S3}$, respectively. We notice that the deviations of the neutrino mass matrix $\Delta M_\nu^{}$ can also be expressed in terms of the parameters $\widehat{A}_i^{}$'s, except that the overall breaking parameters are modified to be $\mathcal{B}_{22}^{} = b/\mathrm{det}(M_R^{})$ and $\mathcal{B}_{32}^{} = c^*/\mathrm{det}(M_R^{})$ for $\mathbf{S2}$ and $\mathbf{S3}$, respectively. We also observe that the analytic expressions for $\Delta \theta_{ij}^{}$'s and $m_i^{}$'s in $\mathbf{S2}$ and $\mathbf{S3}$ are quite similar, and in both scenarios there is no suppression factor of $\theta_{13}^{}$ in $\Delta \theta_{23}^{}$. As a result, one may expect larger deviation in $\theta_{23}^{}$ in $\mathbf{S2}$ and $\mathbf{S3}$ than that in $\mathbf{S1}$. Thus, the last two breaking patterns may be distinguishable from the first one via the future precision measurement of $\theta_{23}^{}$. 

Having discussed some analytical results for the three breaking patterns in $M_D^{}$, we next turn to the detailed numerical analysis. On the one hand, the numerical analysis would extend the analysis to the scenarios where the entries in $\Delta M_\nu^{}$ are not all real. On the other hand, we can also obtain the deviations on the Dirac CP-violating phase $\delta$ and the Majorana phases, which are not easy to obtain analytically. In the numerical analysis, for each breaking pattern we treat all the parameters in $M_D^\prime$ and $M_R^{}$ as free parameters, and vary them within the same ranges as in the previous RG running study. For the breaking parameter $\epsilon$, we vary it as $\epsilon \in [-1, 1]$. The package $\texttt{Multinest}$ is again employed to guide the parameter scan, and the same $\chi^2$ function as before is utilized. 

\begin{figure}
\centering
\includegraphics[scale=0.45]{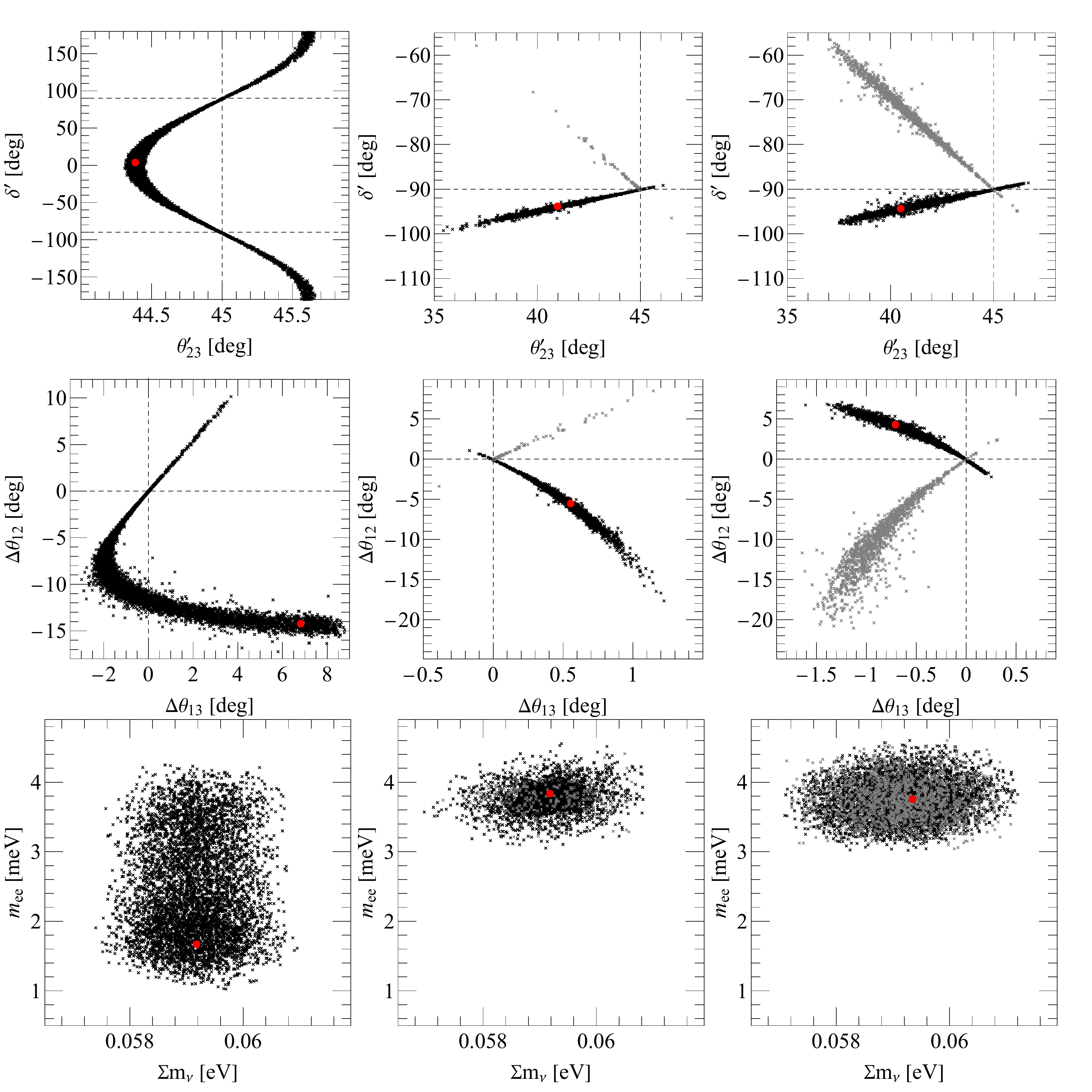}
\caption{\footnotesize Predictions of breaking patterns $\mathbf{S1}$ (left), $\mathbf{S2}$ (middle) and $\mathbf{S3}$ (right) in NH. The scatter points that satisfy $\chi^2 < 30$ are presented by black/gray points, among which the BF point is denoted in red. In $\mathbf{S2}$ and $\mathbf{S3}$ we only show the results that have $\delta^\prime < 0$, and two branches of predictions are distinguished by black and gray points.}
\label{fg:breaking_YD_NH}
\end{figure}

In Fig.~\ref{fg:breaking_YD_NH} we show the numerical results in the case of NH for the three breaking patterns discussed above, i.e., $\mathbf{S1}$ (left), $\mathbf{S2}$ (middle) and $\mathbf{S3}$ (right). Black/gray points have $\chi^2_{} < 30$, and the red point in each case still denotes the best-fit scenario. For $\mathbf{S1}$, $\mathbf{S2}$ and $\mathbf{S3}$, we obtain $\chi^2_{\mathrm{min}} = 9.75, 0.03$ and $0.58$, respectively. Moreover, we only show the results that have $\delta^\prime_{} < 0$ in $\mathbf{S2}$ and $\mathbf{S3}$, as the results for $\delta^\prime_{} > 0$ are quite similar except for a sign change in $\delta^\prime$. 
Lastly, we notice that for $\mathbf{S2}$ and $\mathbf{S3}$ there exist two branches of predictions, which are distinguished by black and gray points. From Fig.~\ref{fg:breaking_YD_NH} we then observe:
\begin{itemize}

\item According to the top three plots, we find that $\Delta \theta_{23}^{}$ in $\mathbf{S1}$ is less than one degree, much smaller than that in the other two breaking patterns. This numerical finding agrees with the analytical results in Table~\ref{tab:break_YD}, i.e., $\Delta \theta_{23}^{}$ is suppressed by a factor of $\theta_{13}^{}$ in $\mathbf{S1}$. 

\item In all three breaking patterns we observe correlations between $\theta_{23}^\prime$ and $\delta^\prime$. For $\mathbf{S1}$, a ``oscillatory" pattern is identified. In Refs.~\cite{Xing:2014zka,Dev:2017fdz,Joshipura:2018rit}, a similar ``oscillatory" correlation between $\theta_{23}^{}$ and $\delta$ was also obtained under the assumption of partial $\mu-\tau$ symmetry in the lepton mixing matrix. However, the ``oscillatory" pattern observed here differs from that in Refs.~\cite{Xing:2014zka,Dev:2017fdz,Joshipura:2018rit} in $\Delta \theta_{23}^{}$, i.e., here $|\Delta \theta_{23}^{}| \lesssim 1^\circ$ while $|\Delta \theta_{23}^{}| \gtrsim 5^\circ$ in Refs.~\cite{Xing:2014zka,Dev:2017fdz,Joshipura:2018rit}. In addition, in $\mathbf{S1}$, $\Delta \delta $ and $\Delta \theta_{23}^{}$ seem to have a negative correlation when $\delta^\prime \sim - 90^\circ$, and the deviation in $\delta^\prime$ is much more dramatic than that in $\theta_{23}^\prime$. $\delta^\prime$ can reach 0 or $\pm 180^\circ$, while $\theta_{23}^{}$ only less than one degree away from $45^\circ$. For $\mathbf{S2}$ and $\mathbf{S3}$, however, the differences between $\Delta \delta $ and $\Delta \theta_{23}^{}$ are less dramatic. In both scenarios there exist two branches of predictions that $\Delta \delta$ and $\Delta \theta_{23}^{}$ can have a positive or negative correlation. In the case with positive correlation $\delta^\prime$ and $\theta_{23}^\prime$ deviate by almost the same amount, while $|\Delta \delta|$ is about three times larger than $|\Delta \theta_{23}^{}|$ for the case with negative correlation. 


\item From the plots in the second row of Fig.~\ref{fg:breaking_YD_NH} we find that $|\Delta \theta_{12}^{}|$ is indeed larger than $|\Delta \theta_{13}^{}|$ and $|\Delta \theta_{23}^{}|$, and it can reach around $15^\circ$ for all three breaking patterns. This is also in agreement with the analytical results given in Table~\ref{tab:break_YD}. Moreover, the value of $\Delta \delta_{12}^{} \sim -15^\circ$ ($5^\circ$) indicates that $\theta_{12}^{}$ before breaking can be quite close to $45^\circ$ ($30^\circ$), and this may have interesting implications in the flavor model building with the exact $\mu-\tau$ reflection symmetry at high energies.


\item Lastly, in the bottom row of Fig.~\ref{fg:breaking_YD_NH} we show the results for the total neutrino mass $\sum m_\nu^{} \equiv m_1^{} + m_2^{} + m_3^{}$ and $m_{ee}^{}$ after the breaking. Here  $m_{ee}^{}$ is the (11) element of $M_\nu^\prime$, and it is responsible for the decay rates of neutrinoless double beta-decay modes of various isotopes. As expected, in NH we have $m_1^{} = 0$ and then satisfying the mass-squared differences from neutrino oscillation experiments leads to $\sum m_\nu^{} \sim 0.06~\mathrm{eV}$. Also, because of NH and $m_1^{} = 0$, the predicted $m_{ee}^{}$ is only a few meV's. Such small values of $\sum m_\nu^{}$ and $m_{ee}^{}$ would be hard to probe by upcoming cosmological observations and $0\nu\beta\beta$ experiments~\cite{Chen:2016qcd,Wang:2017cok,Zhang:2017rbg,Zhang:2015uhk,Guo:2017hea,Zhao:2016ecj}, respectively. 

\end{itemize}

\begin{figure}
\centering
\includegraphics[scale=0.45]{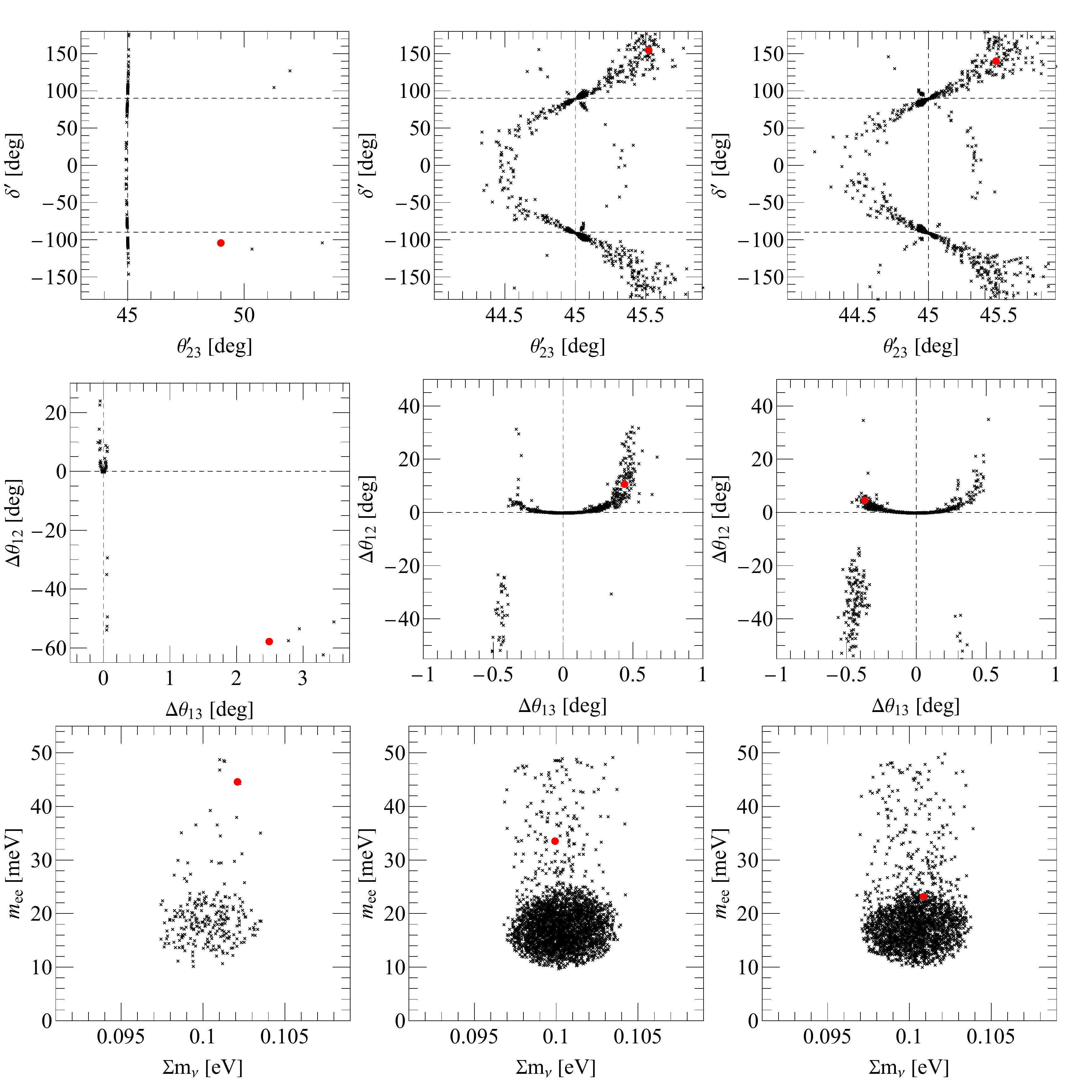}
\caption{\footnotesize Predictions of breaking patterns $\mathbf{S1}$ (left), $\mathbf{S2}$ (middle) and $\mathbf{S3}$ (right) in IH. The scatter points that satisfy $\chi^2 < 50$ are presented by black points, among which the BF point is denoted in red. }
\label{fg:breaking_YD_IH}
\end{figure}

The numerical results for $\mathbf{S1}$, $\mathbf{S2}$ and $\mathbf{S3}$ in the case of IH are shown in Fig.~\ref{fg:breaking_YD_IH}.  Because the deviation in $\theta_{12}^{}$ now has an enhancement factor of $1/\xi \sim 30$, the numerical program is very sensitive to the initial value of $\theta_{12}^{}$ before breaking, so that locating the favored parameter space becomes challenging. The lowest values of $\chi^2_{\mathrm{min}}$ that we are able to obtain are 28.84, 25.75 and 26.19 for $\mathbf{S1}$, $\mathbf{S2}$ and $\mathbf{S3}$, respectively, and the corresponding scatter points are shown in red in Fig.~\ref{fg:breaking_YD_IH}. Moreover, in order to show larger region of parameter space, in this case of IH we require all scatter points (black) to satisfy $\chi^2 < 50$. By inspecting the patterns in Fig.~\ref{fg:breaking_YD_IH}, we then observe:

\begin{itemize}

\item In $\mathbf{S1}$, $\theta_{23}^\prime$ tends to be very close to $45^\circ$, although the BF point has a large deviation in $\theta_{23}^{}$, which may originate from some special combination in the input parameters. The obtained $\delta^\prime$, however, has a large spread in $[-180^\circ, 180^\circ)$. Regarding $\theta_{12}^{}$ and $\theta_{13}^{}$, the deviation in $\theta_{13}^{}$ is quite small in general, while for $\theta_{12}^{}$ large deviations of $\mathcal{O}(10^\circ)$ can be easily achieved. 

\item As for $\mathbf{S2}$ and $\mathbf{S3}$, the favored parameter space are almost the same. Interestingly, in both scenarios it seems that $\delta^\prime$ and $\theta_{23}^\prime$ exhibit similar oscillatory patterns as in the case of $\mathbf{S1}$ under NH. Unfortunately, it is analytically difficult to confirm if there indeed exist connections among these scenarios, especially two different mass orderings are involved. In contrast with $\mathbf{S2}$ and $\mathbf{S3}$ in NH,  the favored $\theta_{23}^\prime$'s are now close to $45^\circ$, while large spreads are observed in $\delta^\prime$. 
On the other hand, the deviations in $\theta_{13}^{}$ are less than one degree in both scenarios, while, as expected, $\theta_{12}^{}$ can easily achieve $\mathcal{O}(10^\circ)$ deviations, due to the enhancement factor of $1/\xi$. 

\item Lastly, in the $m_\mathrm{ee}^{}$ vs. $\sum m_\nu^{}$ plots we observe that for all three breaking scenarios the obtained $\sum m_\nu^{}$'s are close to $0.1~\mathrm{eV}$. This finding agrees with our expectation that with $m_3^{} =0 $ the other neutrino masses $m_1^{}$ and $m_2^{}$ need to be $m_{1,2}^{} \sim 0.05~\mathrm{eV}$ so as to satisfy the currently measured mass-squared differences. For $m_\mathrm{ee}^{}$, although there exist some spread within $[10, 50]~\mathrm{meV}$, most of scatter points are located around $15~\mathrm{meV}$. This is due to the fact that even with breaking the favored $\sigma^\prime$'s after breaking are also quite close to $90^\circ$. As a result, $m_\mathrm{ee}^{}$ can approximate to $m_\mathrm{ee}^{} \sim m_1^{} \cos^2\theta_{12} + e^{2i\sigma} m_2^{} \sin^2\theta_{12} \sim m_1^{} \cos^2\theta_{12} - m_2^{} \sin^2\theta_{12}$, then with $m_1^{} \sim m_2^{}$ we have a significant cancellation between the two terms in $m_\mathrm{ee}^{}$. Therefore, comparing with the NH case, although now we can have larger values of $m_\mathrm{ee}^{}$'s, the value of $\sigma^\prime \sim 90^\circ$ still results in relatively small values of $m_\mathrm{ee}^{} \sim 15~\mathrm{meV}$. Such small values of $m_\mathrm{ee}^{} $ are close to the lower bound of $m_\mathrm{ee}^{}$ in IH, and thus future ton-scale $0\nu\beta\beta$ experiments are needed in order to fully cover the favored parameter space. 

\end{itemize}

So far we have focused on the breaking parameters reside in the neutrino Dirac mass matrix $M_D^{}$. Next, we turn to the breaking patterns in the Majorana mass matrix $M_R^{}$ for the right-handed neutrinos.

\subsection{Breaking $\mu-\tau$ reflection symmetry in $M_R^{}$}

Without loss of generality, we consider two possible breaking patterns in $M_R^{}$. The first breaking pattern arises when $m_{33}^{} = m^{*}_{22} (1+ \epsilon) $, resulting in the Majorana mass matrix $M_R^\prime$ after breaking as,
\begin{eqnarray} \label{eq:S4}
\mathbf{S4:}\quad M_R^\prime = \begin{pmatrix}
m_{22}^{} & m_{23}^{} \\
m_{23}^{} & m^{*}_{22} (1+ \epsilon)
\end{pmatrix} \; ,
\end{eqnarray}
where we have named this breaking scenario as $\mathbf{S4}$. To obtain the other breaking pattern in $M_R^{}$, we exploit the fact that $m_{23}^{}$ needs to be real in the exact $\mu-\tau$ reflection symmetric limit. Assigning some non-zero phase in $m_{23}^{}$ then leads to the other breaking scenario $\mathbf{S5}$, 
\begin{eqnarray} \label{eq:S5}
\mathbf{S5:}\quad M_R^\prime = \begin{pmatrix}
m_{22}^{} & m_{23}^{} e^{i \epsilon \pi} \\
m_{23}^{} e^{i \epsilon \pi} & m_{22}^*
\end{pmatrix} \; .
\end{eqnarray}
Unlike the previous breaking patterns in $M_D^{}$, introducing breaking effects in $M_R^{}$ causes all entries in $\Delta M_\nu^{}$ to be non-zero. For example, for $\mathbf{S4}$ the obtained $\Delta M_\nu^{}$ is given by
\begin{eqnarray}\label{eq:refle_mat_per-mr}
\Delta M_{\nu}^{} \simeq \epsilon \mathcal{B}_{22}^R \begin{pmatrix}
\widehat{A}_1^2 & \widehat{A}_1^{} \widehat{A}_2^{} & \widehat{A}_1^{} \widehat{A}_3^{} \\
\widehat{A}_1^{} \widehat{A}_2^{} & \widehat{A}_2^2 & \widehat{A}_2^{} \widehat{A}_3^{} \\
\widehat{A}_1^{} \widehat{A}_3^{} & \widehat{A}_2^{} \widehat{A}_3^{} & \widehat{A}_3^2
\end{pmatrix}\; ,
\end{eqnarray} 
where $\mathcal{B}_{22}^R$ is defined as $ \mathcal{B}_{22}^R =m^{*}_{22} / \left[\mathrm{det}(M_R^{})\right]^2$, and all $\widehat{A}_i^{}$'s are given in Eq.~(\ref{eq:Aheads}). With all non-zero entries in $\Delta M_\nu^{}$ it is difficult to investigate the breaking effects on neutrino masses and lepton mixing angles analytically. Thus, we employ a similar numerical analysis as the previous breaking scenarios, and both the varied ranges of input parameters and the defined $\chi^2$ function are kept to be the same. 

\begin{figure}
\centering
\includegraphics[scale=0.45]{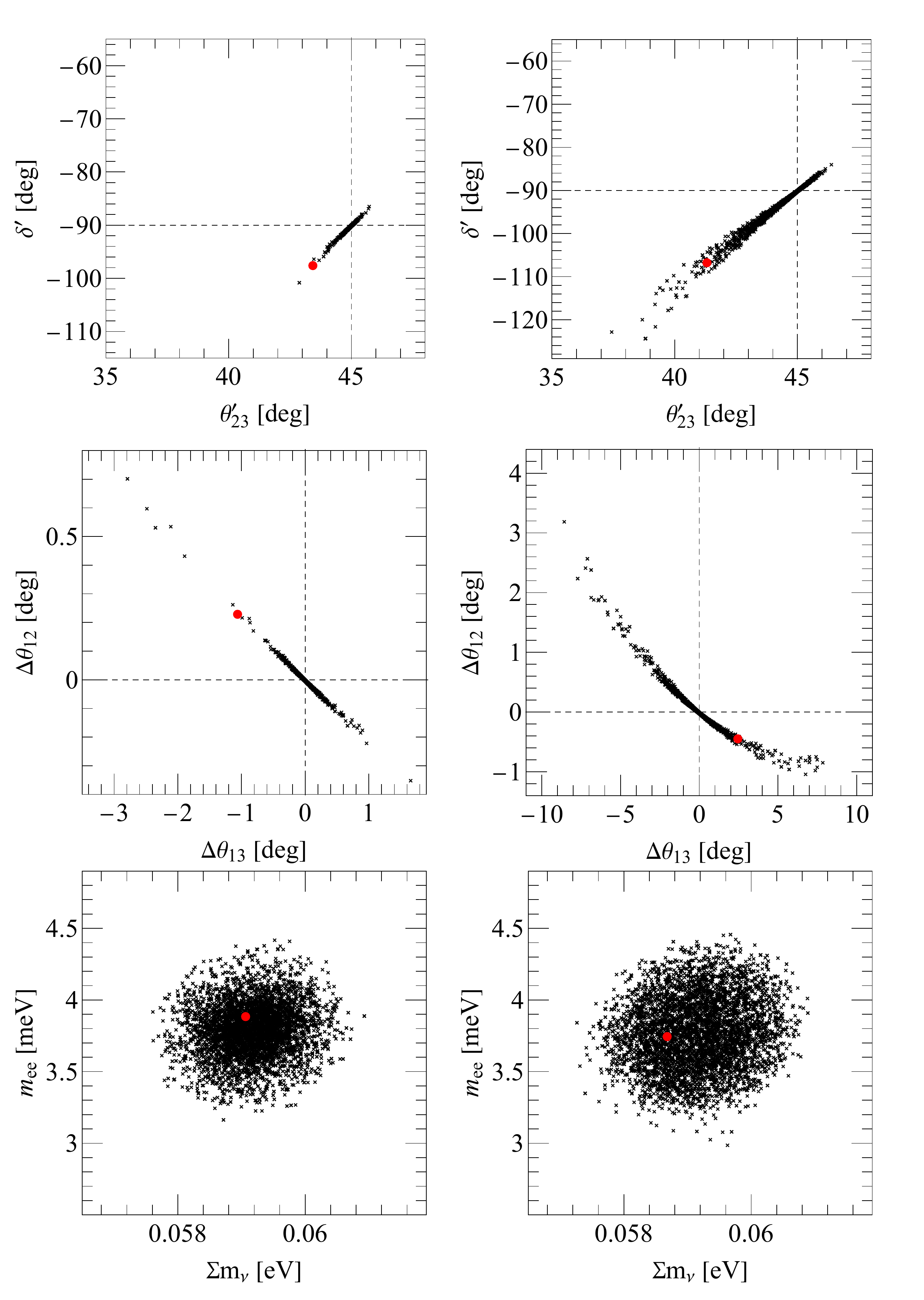}
\caption{\footnotesize Predictions of breaking patterns $\mathbf{S4}$ (left) and $\mathbf{S5}$ (right) in NH. The scatter points that satisfy $\chi^2 < 30$ are presented by black points, among which the BF point is denoted in red. Note that in both scenarios only the results that have $\delta^\prime < 0$ are shown.}
\label{fg:breaking_MR_NH}
\end{figure}

In Fig.~\ref{fg:breaking_MR_NH} we show the numerical results for $\mathbf{S4}$ and $\mathbf{S5}$ in NH. The obtained lowest values of $\chi_\mathrm{min}^2$ are 9.12 and 1.75 for $\mathbf{S4}$ and $\mathbf{S5}$, respectively. Again, in both scenarios only the results that have $\delta^\prime < 0$ are shown, as the other case of $\delta^\prime > 0$ is quite similar except for a sign change in $\delta^\prime$. From Fig.~\ref{fg:breaking_MR_NH}, we first notice that the patterns of the favored parameter space in $\mathbf{S4}$ and $\mathbf{S5}$ are quite similar, although in the latter case more extended parameter space is observed. Between $\theta_{23}^\prime$ and $\delta^\prime$ positive correlations are identified when $\delta^\prime \sim -90^\circ$, and $|\Delta \theta_{23}^{}|$ is about four times smaller than $|\Delta \delta|$. Such a positive correlation of $ \Delta \delta \sim 4 \Delta \theta_{23}^{} $ is different from that in $\mathbf{S2}$ and $\mathbf{S3}$, where the positive correlation gives $ \Delta \delta \sim \Delta \theta_{23}^{} $, so that we may distinguish the breaking scenarios $\mathbf{S4}/\mathbf{S5}$ from $\mathbf{S2}/\mathbf{S3}$ by precisely measuring the correlation between $\theta_{23}^{}$ and $\delta$ in the upcoming experiments. Regarding $\theta_{12}^{}$ and $\theta_{13}^{}$, it seems that $\Delta \theta_{13}^{}$ is negatively correlated to $\Delta \theta_{12}^{}$, and $|\Delta \theta_{13}^{}|$ is about five times larger than $|\Delta \theta_{12}^{}|$. Lastly, as expected, the favored $\sum m_\nu^{}$ is around $0.06~\mathrm{eV}$, and because the Majorana phase $\sigma$ after breaking is still close to $ 90^\circ$, we also have $m_\mathrm{ee}^{} \sim 4~\mathrm{meV}$ as in $\mathbf{S2}$ and $\mathbf{S3}$ under NH. 

\begin{figure}
\centering
\includegraphics[scale=0.45]{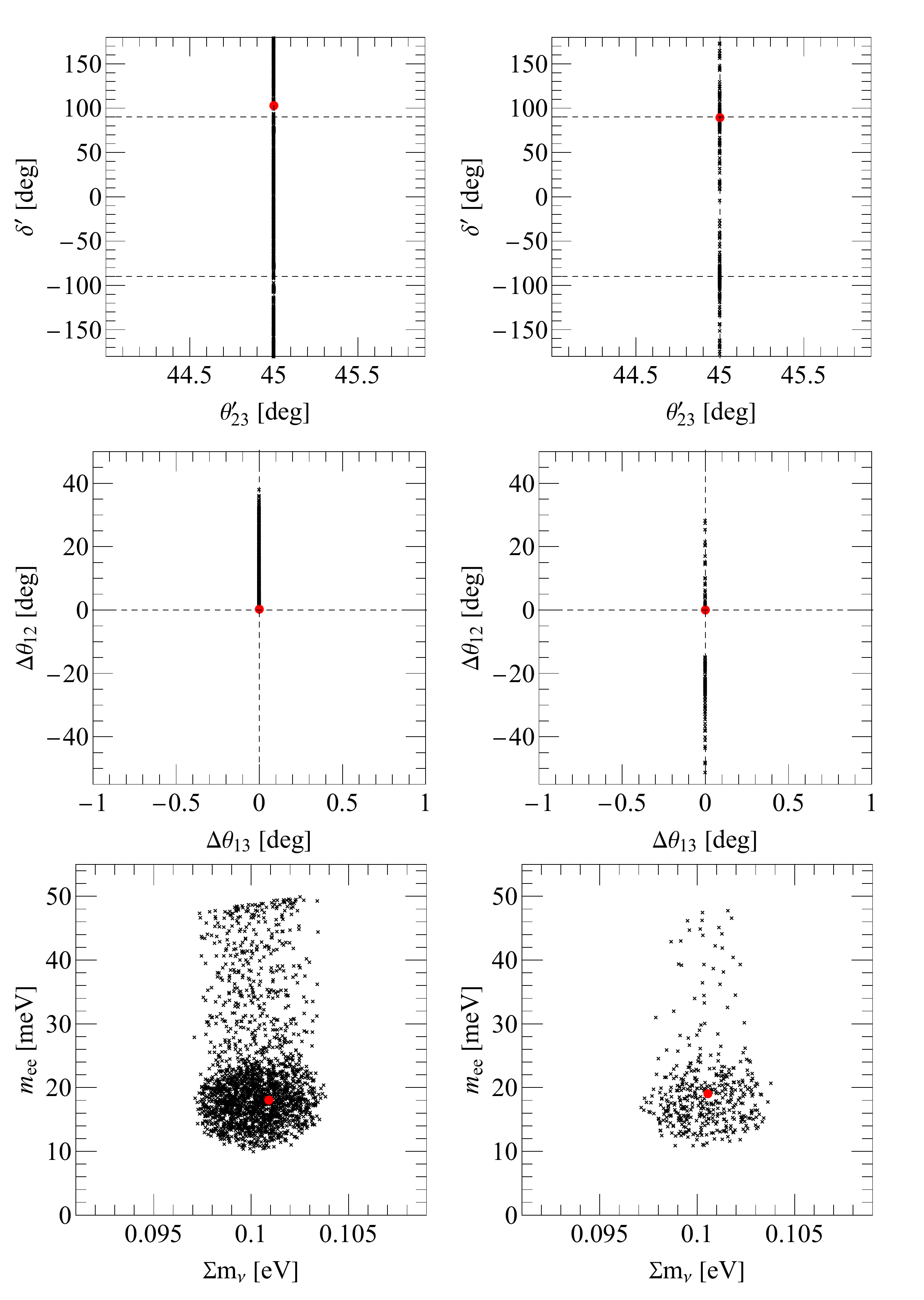}
\caption{\footnotesize Predictions of breaking patterns $\mathbf{S4}$ (left) and $\mathbf{S5}$ (right) in IH. The scatter points that satisfy $\chi^2 < 50$ are presented by black points, among which the BF point is denoted in red. }
\label{fg:breaking_MR_IH}
\end{figure}

We now turn to the numerical results for $\mathbf{S4}$ and $\mathbf{S5}$ in IH, see Fig.~\ref{fg:breaking_MR_IH}. Unexpectedly, we observe that in both $\mathbf{S4}$ and  $\mathbf{S5}$ the deviations in $\theta_{23}^{}$ and $\theta_{13}^{}$ are exactly zero.  A full understanding of such null deviations is hard to pursue by considering a generic form of $M_D^{}$ and $M_R^{}$, while in Appendix~\ref{app:no_dev_IH} we demonstrate such null deviations considering a special case. 
The deviations in $\delta$ and $\theta_{12}^{}$, however, can be rather large. Similar to $\mathbf{S2}/\mathbf{S3}$ in IH, we also have $\sum m_\nu^{} \sim 0.1~\mathrm{eV}$, and because the Majorana phase $\sigma^\prime$ still favors $90^\circ$ after breaking, the preferred $m_\mathrm{ee}^{}$ is again around $15~\mathrm{meV}$.


\begin{table}[!h]
\centering
\begin{tabular}{c|c|c|c|c|c|c}
 \hline
 \hline
 Breaking & $ \theta_{23}^{\prime} $ & $ \delta_{CP}^{\prime}$ &  $ \Delta \theta_{12}^{\prime} $&  $ \Delta \theta_{13}^{\prime} $ & $ \sum m_\nu^{} $ & $ m_{ee} $ \\ 
Scenarios & [deg] & [deg] & [deg] & [deg] & [eV] & [meV] \\
\hline   
$\mathbf{S1}$ & $44.3 \rightarrow 45.7$ & $- 180 \rightarrow 180$ & $- 15 \rightarrow 10$ & $- 1 \rightarrow 9$& $0.0575 \rightarrow 0.061$ & $1 \rightarrow 4.2$    \\ 
\hline
$\mathbf{S2}$ & $35 \rightarrow 46$ & $- 100 \rightarrow - 88$ & $- 18 \rightarrow 1$ & $- 0.1 \rightarrow 1.3$ & $0.057 \rightarrow 0.061$ & $3 \rightarrow 4.5$ \\
              & $40 \rightarrow 45$ & $- 90 \rightarrow - 70$  & $0 \rightarrow 9$ & $0 \rightarrow 1.2$ & --  & -- \\
\hline
$\mathbf{S3}$ & $37.5 \rightarrow 47$ & $- 98 \rightarrow - 88$ & $2 \rightarrow 7$ & $- 1.4 \rightarrow 0.2$ & $0.057 \rightarrow 0.0615$ & $3 \rightarrow 4.5$ \\
              & $46 \rightarrow 47$ & $- 94 \rightarrow - 56$  & $-20  \rightarrow 3$ & $ - 1.7 \rightarrow 0.3$ & --  & -- \\
\hline
$\mathbf{S4}$ & $43 \rightarrow 46$ & $- 100 \rightarrow - 88$ & $- 0.2 \rightarrow 0.7$ & $- 3 \rightarrow 1$ & $0.0575 \rightarrow 0.061$ & $3.1 \rightarrow  4.4$    \\ 
\hline
$\mathbf{S5}$ & $39  \rightarrow 46.5$ & $- 120 \rightarrow -84$ & $- 1 \rightarrow 2.6$ & $- 8 \rightarrow 8$ & $0.057 \rightarrow 0.061$ & $3 \rightarrow 4.5$    \\ 
\hline
\hline
\end{tabular}
\caption{Summary of various breaking scenarios in NH. Note that for $\mathbf{S2}$ and $\mathbf{S3}$ two rows correspond to black and gray patterns in Fig.~\ref{fg:breaking_YD_NH}, respectively.}
\label{tab:break_res_NH}
\end{table}

\begin{table}
\centering
\begin{tabular}{c|c|c|c|c|c|c}
 \hline
 \hline
 Breaking & $ \theta_{23}^{\prime} $ & $ \delta_{CP}^{\prime}$ &  $ \Delta \theta_{12}^{\prime} $&  $ \Delta \theta_{13}^{\prime} $ & $ \sum m_\nu^{} $ & $ m_{ee} $ \\ 
Scenarios & [deg] & [deg] & [deg] & [deg] & [eV] & [meV] \\
\hline   
$\mathbf{S1}$ & $ \sim 45$ & $- 180 \rightarrow 180$ & $0 \rightarrow 20$ & $ \sim 0$ & $0.097 \rightarrow 0.104$ & $10 \rightarrow 50$    \\ 
\hline
$\mathbf{S2}$ & $44.4 \rightarrow 45.7$ & $- 180 \rightarrow 180$ & $- 60 \rightarrow 40$ & $- 0.5 \rightarrow 0.5$ & $0.097 \rightarrow 0.104$ &  $10 \rightarrow 50$ \\
\hline
$\mathbf{S3}$ & $44.4 \rightarrow 45.8$ & $- 180 \rightarrow 180$ & $- 60 \rightarrow 20$ & $- 0.5 \rightarrow 0.5$ & $0.097 \rightarrow 0.104$ &  $10 \rightarrow 50$ \\
\hline
$\mathbf{S4}$ &  $ \sim 45 $ & $- 180 \rightarrow 180$ & $0 \rightarrow 40$ &  $ \sim 0$  & $0.097 \rightarrow 0.104$ &  $10 \rightarrow 50$   \\ 
\hline
$\mathbf{S5}$ & $ \sim 45$ & $- 180 \rightarrow 180$ & $- 30 \rightarrow  25$ & $ \sim 0$  & $0.097 \rightarrow 0.104$ &  $10 \rightarrow 48$   \\ 
\hline
\hline
\end{tabular}
\caption{Summary of various breaking scenarios in IH. }
\label{tab:break_res_IH}
\end{table}

Above we have discussed various breaking scenarios of exact $\mu-\tau$ reflection symmetry, and finally we summarize these results in Table \ref{tab:break_res_NH} and \ref{tab:break_res_IH}. In the context of ongoing neutrino oscillation experiments some of the breaking scenarios can be ruled out. For example, as the latest results of both T2K \citep{Abe:2017uxa} and NO$ \nu $A \cite{NOVA2018} favor NH over IH, and if it remains true, all the breaking schemes corresponding to IH can be ruled out. Furthermore, if in the upcoming experiments $\theta_{23}^{}$ were found to be close to $45^\circ$ within only one degree, the breaking scenarios {\bf S2}, {\bf S3}, {\bf S4} and {\bf S5} in NH would be disfavored. However, to fully exclude these scenarios precise measurement of $\delta_{CP}^{}$ in the future experiments, such as DUNE~\cite{Acciarri:2015uup}, T2HK~\cite{Abe:2014oxa} and MOMENT~\cite{Cao:2014bea}, may be needed.
\section{Conclusion}\label{ses:conc}

In this work we explore the possibility of embedding the $\mu-\tau$ reflection symmetry in the minimal seesaw formalism, where two right-handed neutrinos are added to the SM. Different from the previous works, we apply the $\mu-\tau$ reflection symmetry transformations to both the left- and right-handed neutrinos, resulting in some particular forms of neutrino Dirac mass matrix $M_D^{}$ and the Majorana mass matrix $M_R^{}$ for the right-handed neutrinos. The obtained light neutrino mass matrix $M_\nu^{}$ is found to still possess the usual $\mu-\tau$ reflection symmetry, which predicts maximal atmospheric mixing angle ($ \theta_{23} = 45^\circ$) and Dirac CP phase  ($ \delta = \pm 90 ^\circ$) along with the trivial Majorana phases. We later extend our study by incorporating the breaking of such symmetry, keeping in mind that theoretical as well as experimental results may favor non-maximal $ \theta_{23}$.

The first possible breaking of the symmetry is due to the renormalization group running. Here we choose the minimal supersymmetric standard model as our UV framework, and assume the symmetry to be exact at the GUT scale. When running towards low energies, we encounter three stages of running: above the two seesaw thresholds, between the thresholds, and lastly below the thresholds. Some noteworthy outcomes of our numerical RG analysis are summarized as follows: 
\begin{itemize}
\item The RG running between the thresholds is insignificant, as the two seesaw mass thresholds are found to be quite close. Such closeness of two thresholds is due to the fact that the two columns of the neutrino Yukawa matrix are related by the $\mu-\tau$ reflection symmetry, particularly the symmetry on the right-handed neutrinos as proposed here. 
\item For both NH and IH scenarios, we find that the RG running effects above the seesaw thresholds are comparable to those below the thresholds. This would raise the necessity of considering RG running above the seesaw thresholds, if some flavor symmetry were imposed on the right-handed neutrino fields. 

\item For the three mixing angles, the deviations due to the RG running are all rather small, e.g., $\Delta \theta_{23}^{} \lesssim 1^\circ$, except that for $\theta_{12}^{}$ in IH can there exist a large deviation. The latter exception arises from the fact that the two light neutrino masses may cross each other, leading to an interchange of the order of two neutrino masses. 

\item The RG running effects of the Dirac and Majorana phases are also quite mild in NH, while large deviations of $\mathcal{O}(10^\circ)$ can be observed in the case of IH. 
\item Lastly, we note that the known correlation between the positive/negative deviation of $\theta_{23}^{}$ and the neutrino mass hierarchies are again observed in this extended RG running above the seesaw thresholds.  
\end{itemize}

Having shown that the RG running effects are quite mild, we then proceed to introduce explicit breaking terms in $M_D^{}$ and $M_R^{}$, aiming at obtaining large deviations from the predictions of the exact $\mu-\tau$ reflection symmetry. In total, we systematically investigate five possible breaking patterns, namely,  assigning breaking terms in the (12), (22) and (32) positions of $ M_D^{}$ (denoted as $ \mathbf{S1}$, $ \mathbf{S2}$ and $ \mathbf{S3}$ breaking scenarios) and the (12) and (22) positions of $M_R^{}$ (denoted as $ \mathbf{S4}$ and $ \mathbf{S5}$ breaking scenarios). Both analytical and numerical studies are pursued for $ \mathbf{S1}$, $ \mathbf{S2}$ and $ \mathbf{S3}$, while for $\mathbf{S4}$ and $ \mathbf{S5}$ only the numerical results are attainable. The main results of these breaking scenarios are listed as follows:
\begin{itemize}

\item In NH we find that $\Delta \theta_{23}^{} \lesssim 0.5^\circ$ in $\mathbf{S1}$ while $\Delta \theta_{23}^{}$ of a few degrees can be easily observed for the other breaking patterns. On the other hand, in IH all breaking patterns tend to have $\Delta \theta_{23}^{} \lesssim 0.5^\circ$, especially $\Delta \theta_{23}^{} = 0$ seems to hold exactly for $\mathbf{S4}$ and $\mathbf{S5}$. 

\item For the deviations in $\theta_{13}^{}$, we obtain $\Delta \theta_{13}^{} \lesssim 1^\circ$ for $\mathbf{S2}$, $\mathbf{S3}$ and $\mathbf{S4}$ in NH, while for $\mathbf{S1}$ and $\mathbf{S5}$ deviations of a few degrees are possible. However, in the case of IH all breaking patterns tend to have small deviations in $\theta_{13}^{}$, and again $\Delta \theta_{13}^{} = 0$ seems to hold exactly in $\mathbf{S4}$ and $\mathbf{S5}$ as well. 

\item The deviations in $\theta_{12}^{}$ are found to be around $\mathcal{O}(10^\circ)$ in general, except that for $\mathbf{S4}$ and $\mathbf{S5}$ we observe $\Delta \theta_{12}^{} \lesssim 1^\circ$. 

\item For the Dirac CP-violating phase $\delta$, the resultant values after breaking are extended to the whole range of $[-180^\circ, 180^\circ)$ for $\mathbf{S1}$ in NH and all breaking patterns in IH. For $\mathbf{S2}$, $\mathbf{S3}$, $\mathbf{S4}$ and $\mathbf{S5}$ in NH we identify linear correlations between $\delta$ and $\theta_{23}^{}$ when $\delta \sim 90^\circ$. Such correlations may be tested in the upcoming neutrino experiments. 

\item The Majorana phase $\sigma$ after the breaking tends to favor $90^\circ$, which causes the effective neutrino mass $m_{ee}^{}$ to be around $15~\mathrm{meV}$ for IH while only about $4~\mathrm{meV}$ for NH. Such small values of $m_{ee}^{}$ pose challenges for the upcoming $0\nu\beta\beta$ experiments. 

\end{itemize}

Finally, we conclude the paper with the excitement and caution about the $\mu-\tau$ reflection symmetry. Given the current status of neutrino oscillation data, the $\mu-\tau$ reflection symmetry seems to stand out as a compelling reason for the bewildering flavor puzzles in the lepton sector. However, past two decades also witnessed the shift of the prevailing symmetry pattern in the lepton mixing matrix when more accurate neutrino oscillation data stepped in. Thus, along with continuing the pursuit of the implications behind the $\mu-\tau$ reflection symmetry theoretically, one should also pay close attention to the experimental results in the upcoming years, especially from those measuring the value of $\theta_{23}^{}$.

\begin{acknowledgements}
We like to thank Shun Zhou, Guo-yuan Huang, Jing-yu Zhu and Zhen-hua Zhao for useful discussions. The research work of NN and ZZX  were supported in part by the National Natural Science Foundation of China under grant No. 11775231. JZ was supported in part by the China Postdoctoral Science Foundation under Grant No. 2017M610008.
\end{acknowledgements}

\appendix

\section{Predictions of $\mu-\tau$ reflection symmetry in $M_\nu^{}$}\label{app:convention}

In this appendix we provide the detailed derivation of the predictions in Eq.~(\ref{eq:prediction}), assuming that the light neutrino mass matrix $M_\nu^{}$ possesses the $\mu-\tau$ reflection symmetry, i.e., in the form of  Eq.~(\ref{eq:refle_mat}). To start with, we first perform a 2-3 rotation $U_{23}^{}$ on $M_\nu^{}$ so that the resultant mass matrix is real~\cite{Zhou:2014sya}, namely,
\begin{eqnarray}
U_{23}^\dagger M_\nu^{} U_{23}^* = -\begin{pmatrix}
A & \sqrt{2} \mathrm{Im}(B) & \sqrt{2} \mathrm{Re}(B) \\
\sqrt{2} \mathrm{Im}(B) & D - \mathrm{Re}(C) & \mathrm{Im}(C) \\
 \sqrt{2} \mathrm{Re}(B)  & \mathrm{Im}(C) & D +\mathrm{Re}(C)
\end{pmatrix} ,
\end{eqnarray}
with $U_{23}^{}$ given by
\begin{eqnarray}
U_{23}^{} = \begin{pmatrix}
1 & 0 & 0 \\
0 & \frac{i}{\sqrt{2}} & \frac{1}{\sqrt{2}} \\
0 & \frac{-i}{\sqrt{2}} & \frac{1}{\sqrt{2}}
\end{pmatrix} . 
\end{eqnarray}
The above real mass matrix can be further diagonalized by an orthogonal matrix $O$, 
\begin{eqnarray}
O = \begin{pmatrix}
\eta_e^{} & 0 & 0 \\
0 & \eta_\mu^{} & 0 \\
0 & 0 & \eta_\tau^{}
\end{pmatrix}
\begin{pmatrix}
1 & 0 & 0 \\
0 & c_1^{} & s_1^{} \\
0 & - s_1^{} & c_1^{}
\end{pmatrix}
\begin{pmatrix}
c_2^{} & 0 & s_2^{} e^{-i \delta_0^{}} \\
0 & 1 & 0 \\
-s_2^{} e^{i \delta_0^{}} & 0 & c_2^{}
\end{pmatrix}
\begin{pmatrix}
c_3^{} & s_3^{} & 0 \\
-s_3^{} & c_3^{} & 0 \\
0 & 0 & 1
\end{pmatrix} P_R^{}
\end{eqnarray}
where $P_R^{} = \mathrm{diag}\{\eta_\rho^{}, \eta_\sigma^{}, 1\}$, so that 
\begin{eqnarray}
P_M^{} O^T U_{23}^\dagger M_\nu^{} U_{23}^* O P_M^{} =  \mathrm{diag}\{m_1^{}, m_2^{}, m_3^{}\} \; .
\end{eqnarray}
Here $c_i = \cos\theta_i^{}$ and $s_i^{} = \sin\theta_i^{}$ for $i = 1, 2, 3$, and we take $\eta_\alpha^{} (\alpha = e, \mu, \tau, \rho, \sigma) = \pm 1$ and $\delta_0^{} = 0, \pi$ to ensure that all $\theta_i^{}$'s are within $[0, \pi/2)$. For instance, if $\theta_3^{}$ were in the fourth quadrant, one could bring it back to the first quadrant, i.e., $\theta_3^{} \rightarrow - \theta_3^{}$, via the simultaneous transformations of $\eta_e^{} \rightarrow - \eta_e^{}$, $\eta_\rho \rightarrow - \eta_\rho^{}$ and $\delta_0^{} \rightarrow \delta_0^{} + \pi$. In addition, to keep all neutrino masses $m_i^{}$'s to be positive, we introduce a diagonal phase matrix $P_M^{} = \mathrm{diag}\{ \sqrt{\epsilon_\rho}, \sqrt{\epsilon_\sigma}, 1 \}$ with $\epsilon_{\rho,\sigma}^{} = \pm 1$. 

The overall neutrino mixing matrix $V$ can then be read out,
\begin{eqnarray}
V &=& U_{23}^{} O P_M^{} \nonumber \\
& = & \eta_\tau^{} \begin{pmatrix}
\frac{\eta_e^{}}{\eta_\tau^{}} & 0 & 0 \\
0 & e^{i (\theta_1^\prime + \chi)} & 0 \\
0 & 0 & e^{-i \theta_1^\prime} 
\end{pmatrix}
\begin{pmatrix}
1 & 0 & 0 \\
0 & \frac{1}{\sqrt{2}} & \frac{e^{-i \chi}}{\sqrt{2}} \\
0 & \frac{-e^{i\chi}}{\sqrt{2}} & \frac{1}{\sqrt{2}}
\end{pmatrix}
\begin{pmatrix}
c_2^{} & 0 & s_2^{} e^{-i \delta_0^{}} \\
0 & 1 & 0 \\
-s_2^{} e^{i \delta_0^{}} & 0 & c_2^{}
\end{pmatrix}
\begin{pmatrix}
c_3^{} & s_3^{} & 0 \\
-s_3^{} & c_3^{} & 0 \\
0 & 0 & 1
\end{pmatrix} P_R^{} P_M^{} \nonumber \\
\end{eqnarray}
where $\theta_1^\prime = \theta_1^{} \eta_\mu^{} / \eta_\tau^{}$ and $\chi = \mathrm{arg}( i \eta_\mu^{} / \eta_\tau^{}) = \pm \pi/2$. 
The product of three rotation matrices in the above equation are in the form of 
\begin{eqnarray}
U_R^{} = \begin{pmatrix}
1 & 0 & 0 \\
0 & c_{23}^{} & s_{23} e^{-i \delta_{23}^{}} \\
0 & -s_{23} e^{i \delta_{23}^{}} & c_{23}^{}
\end{pmatrix}
\begin{pmatrix}
c_{13}^{} & 0 & s_{13}^{} e^{-i \delta_{13}^{}} \\
0 & 1 & 0 \\
-s_{13}^{} e^{i \delta_{13}^{}} & 0 & c_{13}^{}
\end{pmatrix}
\begin{pmatrix}
c_{12}^{} & s_{12}^{}e^{-i \delta_{12}^{}} & 0 \\
-s_{12}^{}e^{i \delta_{12}^{}} & c_{12}^{} & 0 \\
0 & 0 & 1
\end{pmatrix},
\end{eqnarray}
where $c_{ij}^{} = \cos\theta_{ij}^{}$ and $s_{ij} = \sin\theta_{ij}^{}$. We then have $\theta_{23} = \pi/4$, $\theta_{13}^{} = \theta_2^{}$, $\theta_{12}^{} = \theta_3^{}$, $\delta_{23}^{} = \chi$, $\delta_{13} = \delta_0^{}$ and $\delta_{12}^{} = 0$. 
Furthermore, it is known that $U_R^{}$ can be recasted into a form that is in the PDG convention~\cite{King:2002nf},
\begin{eqnarray}
U_R^{} = \begin{pmatrix}
e^{- i (\delta_{12}^{} + \delta_{23}^{}) } & 0 & 0 \\
0 & e^{-i \delta_{23}^{}} & 0 \\
0 & 0  & 1
\end{pmatrix}
\begin{pmatrix}
1 & 0 & 0 \\
0 & c_{23}^{} & s_{23}\\
0 & -s_{23} & c_{23}^{}
\end{pmatrix}
\begin{pmatrix}
c_{13}^{} & 0 & s_{13}^{} e^{-i \delta} \\
0 & 1 & 0 \\
-s_{13}^{} e^{i \delta} & 0 & c_{13}^{}
\end{pmatrix}
\begin{pmatrix}
c_{12}^{} & s_{12}^{} & 0 \\
-s_{12}^{} & c_{12}^{} & 0 \\
0 & 0 & 1
\end{pmatrix} P_R^\prime \; ,
\end{eqnarray}
where $\delta = \delta_{13}^{} - \delta_{23} - \delta_{12}$ and $P_R^\prime = \mathrm{diag}\{e^{i (\delta_{12}^{} + \delta_{23}^{} )}, e^{i \delta_{23}^{}}, 1 \}$. Applying such a transformation into $V$ then yields
\begin{eqnarray}
V = \eta_\tau^{} \begin{pmatrix}
\frac{\eta_e^{}}{\eta_\tau^{}} e^{- i \delta_{23}^{}} & 0 & 0 \\
0 & e^{i \theta_1^\prime} & 0 \\
0 & 0 & e^{-i \theta_1^\prime} 
\end{pmatrix}
\begin{pmatrix}
1 & 0 & 0 \\
0 & \frac{1}{\sqrt{2}} & \frac{1}{\sqrt{2}} \\
0 & \frac{-1}{\sqrt{2}} & \frac{1}{\sqrt{2}}
\end{pmatrix}
\begin{pmatrix}
c_{2}^{} & 0 & s_{2}^{} e^{-i \delta} \\
0 & 1 & 0 \\
-s_{2}^{} e^{i \delta} & 0 & c_{2}^{}
\end{pmatrix}
\begin{pmatrix}
c_{3}^{} & s_{3}^{} & 0 \\
-s_{3}^{} & c_{3}^{} & 0 \\
0 & 0 & 1
\end{pmatrix} P_\nu^{} \;,
\end{eqnarray}
where $P_\nu^{} = P_R^\prime P_R^{} P_M^{}$. Comparing the above equation with Eq.~(\ref{eq:pmns}) and ignoring the overall phase $\eta_\tau^{}$, we obtain 
\begin{eqnarray}
\theta_{12} &=& \theta_3, \quad
\theta_{13} =  \theta_2, \quad
\theta_{23} = \pi/4, \quad
 \delta = \delta_0^{} - \delta_{23}^{} = \pm \pi/2, \\
\rho &=& \mathrm{arg}( \eta_\rho^{} \sqrt{\epsilon_\rho^{}} e^{i \delta_{23}} ) = 0, \pi, \pm \pi/2, \quad
\sigma = \mathrm{arg}(\eta_\sigma \sqrt{\epsilon_\sigma^{}} e^{i \delta_{23}} ) = 0, \pi, \pm \pi/2, \\
\phi_e^{} &=& \mathrm{arg}(\eta_e^{} e^{- i \delta_{23}^{}}/ \eta_\tau^{}) = \pm \pi/2, \qquad ~
 \phi_\mu^{} = - \phi_\tau = \theta_1^\prime = \pm \theta_1^{} \; .
\end{eqnarray}
Note that taking $\phi_e^{} \rightarrow \phi_e^{} + \pi$, $\phi_\mu^{} \rightarrow \phi_\mu^{} + \pi$ and $\phi_\tau^{} \rightarrow \phi_\tau^{} - \pi$ only changes the overall sign of $V$, while still maintaining the relation $\phi_\mu^{} = - \phi_\tau^{}$. Thus, $\phi_e^{}$ can be restricted to $\phi_e^{} = \pi/2$. Moreover, for $\rho$ and $\sigma$ we can also have $\rho \rightarrow \rho + \pi$ and $\sigma \rightarrow \sigma + \pi$ without modifying $M_\nu^{}$, and therefore we obtain $\rho, \sigma = 0$ or $\pi/2$. It is worth pointing out that if $\theta_{23}^{}$ were chosen to be $\theta_{23}^{} = -\pi/4$, $\delta_{23}^{}$ would then be $\chi + \pi$, and in that case in order to keep the relation $\phi_\mu^{} = - \phi_\tau^{}$, we would have $\phi_e^{} = 0$ after separating out some overall phases.

\section{$\chi^2$ function}\label{app:chi_sq}

Here we define the Gaussian-$\chi^2$ function that has been adopted in numerical analysis as,
\begin{equation}
\chi^{2} = \sum_i \dfrac{\left[  \xi_i^{\rm true} - \xi_i^{\rm test} \right] ^{2}  }{\sigma \left[ \xi_i^{\rm true} \right] ^{2}} \;,
\end{equation}
where $\xi$ represents the neutrino oscillation parameters, i.e., $\xi = \{ \Delta m_{21}^{2}, |\Delta m_{31}^{2}|,  \theta_{12}, \theta_{13} , \theta_{23} \}$. $\xi_i^{\mathrm{ture}}$'s represent the best-fit values from the recent global fit results~\cite{deSalas:2017kay}, while $\xi_i^{\mathrm{test}}$'s are the predicted values for a given set of parameters in theory. Note that for $\sigma \left[ \xi_i^{\rm true} \right]$ we symmetrize the 1-$\sigma$ errors given in Ref.~\cite{deSalas:2017kay}. 

\section{Null deviations in $\theta_{23}^{}$ and $\theta_{13}^{}$ for $\mathbf{S4}$ and $\mathbf{S5}$ in IH}\label{app:no_dev_IH}

We here choose special forms of $M_D^{}$ and $M_R^{}$ in $\mathbf{S4}$ to demonstrate that there exist no deviations in $\theta_{23}^{}$ and $\theta_{13}^{}$ in IH.  Similarly, one can apply the following discussion to $\mathbf{S5}$, where the same conclusions hold.  The special forms of $M_D^{}$ and $M_R^{}$ are respectively given by
\begin{eqnarray}
M_D^{} = \begin{pmatrix}
i b & - i b \\
c & d \\
d & c
\end{pmatrix}, \quad M_R^{} = \begin{pmatrix}
m_{22}^{} & 0 \\
0 & m_{22}(1+\epsilon)
\end{pmatrix} \;, 
\end{eqnarray}
where $b, c, d, m_{22}^{}$ and $\epsilon$ are all real. From the seesaw formula, we obtain the light neutrino mass matrix $M_\nu^\prime$ as
\begin{eqnarray}
M_\nu^\prime & =&  - M_D^{} M_R^{-1} M_D^T \nonumber \\
& = &  -\frac{1}{m_{22}^{}(1+\epsilon)}\begin{pmatrix}
-2 b^2 - b^2 \epsilon & i b (c-d) + i b c \epsilon & - i b (c-d) + i b d \epsilon \\
i b (c-d) + i b c \epsilon & c^2 + d^2 + c^2 \epsilon & 2 c d + c d \epsilon \\
 - i b (c-d) - i b c \epsilon & 2 c d + c d \epsilon & c^2 + d^2 + d^2 \epsilon
\end{pmatrix} \; .
\end{eqnarray}
It is apparent that with $\epsilon \rightarrow 0$, the above $M_\nu^\prime$ possesses the exact $\mu-\tau$ reflection symmetry. Next, we first perform a (23) rotation $R_{23}^{}$ of $\pi/4$ on $M_\nu^\prime$, namely,
\begin{eqnarray}
R_{23}^T M_\nu^\prime R_{23}^{} = -\frac{1}{2 m_{22}^{}(1+\epsilon)} \begin{pmatrix}
- 2 b^2( \epsilon + 2) & i \sqrt{2} b (c+d)\epsilon & - i \sqrt{2}b (c-d) (\epsilon +2 ) \\
i \sqrt{2} b (c+d)\epsilon & (c+d)^2 (\epsilon + 2) & (d^2 - c^2) \epsilon \\
- i \sqrt{2}b (c-d) (\epsilon +2 ) & (d^2 - c^2) \epsilon & (c -d )^2 (\epsilon + 2) 
\end{pmatrix} \; , \nonumber
\end{eqnarray}
with $R_{23}^{}$ given by
\begin{eqnarray}
R_{23}^{} = \begin{pmatrix}
1 & 0 & 0 \\
0 & \frac{1}{\sqrt{2}} & -  \frac{1}{\sqrt{2}}\\
0 &  \frac{1}{\sqrt{2}} &  \frac{1}{\sqrt{2}}
\end{pmatrix} \; .
\end{eqnarray}
The phase in $R_{23}^T M_\nu^\prime R_{23}^{}$ can be further removed by a diagonal phase matrix $P_\phi^{} = \mathrm{diag}\{i, 1, 1\}$, resulting in a pure real mass matrix as follows,
\begin{eqnarray}
P_\phi^{} R_{23}^T M_\nu^\prime R_{23}^{} P_\phi^{} = -\frac{1}{2 m_{22}^{}(1+\epsilon)} \begin{pmatrix}
- 2 b^2( \epsilon + 2) & \sqrt{2} b (c+d)\epsilon & - \sqrt{2}b (c-d) (\epsilon +2 ) \\
\sqrt{2} b (c+d)\epsilon & (c+d)^2 (\epsilon + 2) & (d^2 - c^2) \epsilon \\
- \sqrt{2}b (c-d) (\epsilon +2 ) & (d^2 - c^2) \epsilon & (c -d )^2 (\epsilon + 2) 
\end{pmatrix} \; . \nonumber
\end{eqnarray}
Surprisingly, we then notice that the above matrix can be brought in a block diagonal form with a (13) rotation $R_{13}^{}$, whose mixing angle $\theta_{13}^{}$ coincides with the case without breaking! To be explicit, $R_{13}^{}$ is given by
\begin{eqnarray}
R_{13}^{} = \begin{pmatrix}
\cos\theta_{13}^{} & 0 & \sin\theta_{13}^{} \\
0 & 1 & 0 \\
- \sin\theta_{13}^{} & 0 & \cos\theta_{13}^{}
\end{pmatrix} \; ,
\end{eqnarray}
where $\theta_{13}^{} = -\frac{1}{2} \tan^{-1} \left[ \frac{2\sqrt{2}b(c-d)}{2 b^2 - (c - d)^2} \right ]$.

Depending on the sign of $2b^2 - (c-d)^2$, we then have two scenarios corresponding to different mass hierarchies. When $2b^2 - (c-d)^2 > 0 $, the resultant matrix after performing the (13) rotation $R_{13}^{}$ is given by
\begin{eqnarray}
R_{13}^T P_\phi^{} R_{23}^T M_\nu^\prime R_{23}^{} P_\phi^{} R_{13}^{} = - \frac{1}{2 m_{22}^{} (1+\epsilon)} \begin{pmatrix}
-  \left[ 2 b^2 + (c-d)^2 \right](\epsilon + 2) & \sqrt{2 b^2 + (c-d)^2} (c+d) \epsilon & 0 \\
\sqrt{2 b^2 + (c-d)^2} (c+d) \epsilon & -(c+d)^2 (\epsilon + 2)  & 0 \\
0 & 0 & 0 
\end{pmatrix} \; . \nonumber
\end{eqnarray}
It is then apparent that the above scenario corresponds to the IH case, as $m_3^{} = 0$, and the above matrix can be finally diagonalized by a (12) rotation. Since in the above diagonalization procedure we follow a (23)-(13)-(12) sequence that is same as the PDG convention, the mixing angles of $\theta_{23}^{}$ and $\theta_{13}^{}$ stay the same as the case without breaking. Note that if $\epsilon = 0$ the above matrix is already diagonalized, and thus we have $\theta_{12}^{} = 0$ without breaking; with breaking we instead require $\theta_{12}^{} \neq 0$, so that $\theta_{12}^{}$ is not immune to the breaking in $M_R^{}$. On the other hand, we arrive at the NH case when $2b^2 - (c-d)^2 < 0 $, namely,
\begin{eqnarray}
R_{13}^T P_\phi^{} R_{23}^T M_\nu^\prime R_{23}^{} P_\phi^{} R_{13}^{} \qquad \qquad \qquad \qquad \qquad \qquad \qquad \qquad \qquad \qquad \qquad \qquad \qquad  \nonumber \\
= - \frac{1}{2 m_{22}^{} (1+\epsilon)} \begin{pmatrix}
0 & 0 & 0 \\
0 & - (c+d)^2 (\epsilon + 2) &  \sqrt{2 b^2 + (c-d)^2} (c+d)~\mathrm{sgn}(c-d) \epsilon \\
0 & \sqrt{2 b^2 + (c-d)^2} (c+d)~\mathrm{sgn}(c-d) \epsilon & -  \left[ 2 b^2 + (c-d)^2 \right](\epsilon + 2) 
\end{pmatrix} \; , \nonumber
\end{eqnarray}
where $\mathrm{sgn}(x)$ stands for the sign of  $x$. Now because we need another (23) rotation, the final (23) rotation angle would deviate from $\pi/4$ when adopting the PDG convention, and in the meantime $\theta_{13}^{}$ would also get modified. As a result, in this NH case we expect that all the three mixing angles can be affected by the breaking terms in $M_R^{}$.

\section{Details of best-fit scenarios in the RG running study}\label{app:rge_table}

In Table~\ref{tab:result_rge} we provide the neutrino Yukawa matrix and the Majorana neutrino mass matrix at the high energy boundary for both the hierarchies.  We also give 
the detailed numerical values of all the neutrino oscillation parameters at the various energy scales.
 
\begin{table}[h!]
\centering
\scriptsize
\begin{tabular}{c | c | c | c | c | c | c | c | c | c | c | c | c }
\hline
\hline
& \multicolumn{4}{c}{NH(BF)} & \multicolumn{4}{|c}{IH(BF)} & \multicolumn{4}{|c}{IH(2)} \\
\hline
$Y_\nu^{} (\Lambda_\mathrm{GUT}^{})$ &  \multicolumn{4}{c}{$\begin{bmatrix}
0.166 + 0.013i & 0.166 - 0.013i \\
0.077 + 0.232i & 0.344 - 0.395i \\
0.344 + 0.395i & 0.077 - 0.232i
\end{bmatrix}$} & \multicolumn{4}{|c}{$\begin{bmatrix}
-0.744 + 0.147i & -0.744 - 0.147i \\
-0.389 - 0.065i & 0.037 + 0.242 i \\
 0.037 - 0.242 i & -0.389 - 0.065i\\
\end{bmatrix}$} & \multicolumn{4}{|c}{$\begin{bmatrix}
0.823 + 0.222i & 0.823 - 0.222i \\
0.173 + 0.447i & -0.608+0.223i \\
-0.608 - 0.223i & 0.173 - 0.447i \\
\end{bmatrix}$} \\
%
%
$\frac{M_R^{} (\Lambda_\mathrm{GUT}^{})}{10^{14}\mathrm{GeV}}$ &  \multicolumn{4}{c}{$\begin{bmatrix}
0.396 + 0.041i & 2.956 \\
2.956 & 0.396 - 0.041i 
\end{bmatrix}$} & \multicolumn{4}{|c}{$\begin{bmatrix}
1.033 + 2.611 i & 0.470 \\
0.470 & 1.033 - 2.611 i
\end{bmatrix}$} & \multicolumn{4}{|c}{$\begin{bmatrix}
5.802 - 1.086i & 2.761 \\
2.761 & 5.802 + 1.086i
\end{bmatrix}$} \\
\hline
$\left(\frac{M_1^{}}{10^{14}\mathrm{GeV}}, \frac{M_2^{}}{10^{14}\mathrm{GeV}} \right )$ &  \multicolumn{4}{c}{(2.438, 3.268)} &  \multicolumn{4}{|c}{(2.187, 3.002)} &  \multicolumn{4}{|c}{(2.932, 7.503)}  \\
\hline
& $~\Lambda_{\mathrm{GUT}}^{}~$ &  $~~~M_2^{}~~$ & $~~~M_1^{}~~$ & $\Lambda_\mathrm{EW}^{}$ & $~\Lambda_{\mathrm{GUT}}^{}~$ &  $~~~M_2^{}~~$ & $~~~M_1^{}~~$ & $\Lambda_\mathrm{EW}^{}$ & $~\Lambda_{\mathrm{GUT}}^{}~$ &  $~~~~M_2^{}~~~$ & $~~~~M_1^{}~~~$ & $\Lambda_\mathrm{EW}^{}$ \\
\hline
$\theta_{12}^{}~[^\circ]$ & 34.893 & 35.000 & 35.000 & 35.130 & 34.664 & 35.014 & 34.995 & 35.229 & 56.640 & 32.720 & 32.810 & 33.092 \\
$\theta_{13}^{}~[^\circ]$ & 8.394 & 8.504 & 8.513 & 8.559  & 8.685 & 8.447 & 8.439 & 8.398   & 9.077 & 8.836 & 8.828 & 8.784  \\
$\theta_{23}^{}~[^\circ]$ & 45 & 45.385 & 45.408 & 45.631 & 45 & 44.700 & 44.695 & 44.417 & 45 & 44.337 & 44.253 & 43.977 \\
$\delta~[^\circ]$ & 90 & 89.992 & 90.010 & 90.376  & $-90$ & $-91.376$ & $-91.334$ & $-103.307$  & $-90$ & 90.432 & 90.960 & 103.176\\
$\sigma~[^\circ]$ & 90 & 90.048 & 90.033 & 90.021 & 90 & 89.575 & 89.588 & 85.600 & 90 & 90.257 & 90.460 & 95.387 \\
$m_1^{}~[10^{-2}~\mathrm{eV}]$ & 0 & 0 & 0 & 0 & 6.820 & 6.147 & 6.123 & 5.067 & 6.910 & 5.953 & 5.917 & 4.946 \\
$m_2^{}~[10^{-2}~\mathrm{eV}]$ & 1.121 & 1.077 & 1.075 & 0.875 & 7.007 & 6.246 & 6.232 & 5.143 & 7.034 & 6.111 & 6.031 & 5.023  \\
$m_3^{}~[10^{-2}~\mathrm{eV}]$ & 6.643 & 6.229 & 6.209 & 5.040 & 0 & 0 & 0 & 0 & 0 & 0 & 0 & 0  \\
\hline
\hline
\end{tabular}
\caption{Details of best-fit scenarios in the RG running study. }
\label{tab:result_rge}
\end{table}

\bibliographystyle{apsrev}
\bibliography{mu-tau}
\end{document}